\documentclass[11pt,a4paper]{article}

\usepackage{placeins}

\usepackage{microtype}
\usepackage{adjustbox}
\usepackage{enumitem}

\usepackage{amsthm} \usepackage{thmtools} 

\theoremstyle{definition}
\newtheorem{theorem}{Theorem}[section]

\newtheorem{lemma}[theorem]{Lemma}

\usepackage{graphicx}
\usepackage{gnuplot-lua-tikz}
\usepackage{url}
\usepackage[caption=false]{subfig}
\usepackage{amssymb}

\usepackage[T1]{fontenc}

\newcommand*{\instancename}[1]{\ensuremath{\mathsf{#1}}} \newcommand*{\functionname}[1]{{\ensuremath{\renewcommand{\rmdefault}{ptm}\fontfamily{ppl}\selectfont\textrm{\textup{#1}}}}} 

\newcommand*{\teigi}[1]{\emph{\color{blue}#1}}

\makeatletter
\newcommand*{\UnaryOperator}{\@ifstar
	\UnaryOperatorS \UnaryOperatorN }
\newcommand*{\Oh}{\@ifstar
	\OhS \OhN }
\makeatother

\newcommand*{\UnaryOperatorS}[2][]{\ifx&#1&\ensuremath{\mathop{}\mathopen{}#2\mathopen{}}\else \ensuremath{\mathop{}\mathopen{}#2\mathopen{}\left(#1\right)}\fi }
\newcommand*{\UnaryOperatorN}[2][]{\ifx&#1&\ensuremath{\mathop{}\mathopen{}#2\mathopen{}}\else \ensuremath{\mathop{}\mathopen{}#2\mathopen{}(#1)}\fi }

\DeclareMathAlphabet{\mathup}{OT1}{\familydefault}{m}{n}
\newcommand*{\bigOh}{\mathcal{O}}
\newcommand*{\OhS}[1]{\UnaryOperator*[#1]{\bigOh}}
\newcommand*{\OhN}[1]{\UnaryOperator[#1]{\bigOh}}

\newcommand*{\Om}[1]{\UnaryOperator[#1]{\mathup{\Omega}}}

\newcommand*{\Ot}[1]{\UnaryOperator[#1]{\mathup{\Theta}}}

\newcommand*{\TriePP}{\mbox{\instancename{c}-\instancename{trie}\kern0.3ex\raisebox{-0.0ex}{\scalebox{0.8}{\kern-0.4ex+}}\kern-0.2ex\raisebox{0.4ex}{\scalebox{0.8}{\kern-0.4ex+}}}}

\newcommand*{\CPlusPlus}{C\kern0.3ex\raisebox{-0.0ex}{\scalebox{0.8}{\kern-0.4ex+}}\kern-0.2ex\raisebox{0.4ex}{\scalebox{0.8}{\kern-0.4ex+}}}

\newcommand*{\DicChild}[1]{\ensuremath{\instancename{DicChild}_{#1}}}
\newcommand*{\DicHandle}{\instancename{DicHandle}}
\newcommand*{\DicChunk}{\instancename{DicChunk}}

\newcommand*{\nodeV}{\ensuremath{v}}
\newcommand*{\nodeU}{\ensuremath{u}}
\newcommand*{\nodeW}{\ensuremath{w}}

\newcommand*{\fnExtent}{\functionname{extent}}
\newcommand*{\fnKey}{\functionname{key}}
\newcommand*{\fnHandle}{\functionname{handle}}
\newcommand*{\fnExit}{\functionname{exit}}
\newcommand*{\fnParex}{\functionname{parex}}
\newcommand*{\fnAccess}{\functionname{access}}

\newcommand*{\setKeys}{\ensuremath{\mathcal{K}}}

\newcommand*{\fnInsert}{\functionname{insert}}
\newcommand*{\fnLookup}{\functionname{lookup}}
\newcommand*{\fnLocatePrefix}{\functionname{locatePrefix}}
\newcommand*{\fnDelete}{\functionname{delete}}

\newcommand*{\triePCTXOR}{\ensuremath{\textsf{PCT}_{\textup{bit}}}}
\newcommand*{\triePCTHash}{\ensuremath{\textsf{PCT}_{\textup{hash}}}}

\begin{document}
\title{c-trie++: A Dynamic Trie Tailored for Fast Prefix Searches\footnote{Parts of this work have already been presented at 
the Data Compression Conference (DCC) 2020~\cite{tsuruta20ctrie}.}}
\date{}

\author{Kazuya~Tsuruta, Dominik~K\"oppl, Shunsuke~Kanda, Yuto~Nakashima, \\ Shunsuke~Inenaga, Hideo~Bannai, Masayuki~Takeda}

\maketitle              

\begin{abstract}
Given a dynamic set \setKeys{} of $k$ strings of total length $n$ whose characters are drawn from an alphabet of size~$\sigma$,
a keyword dictionary is a data structure built on $\setKeys$ that provides lookup, prefix search, and update operations on~$\setKeys$.
Under the assumption that $\alpha = w/ \lg \sigma$ characters fit into a single machine word of~$w$ bits,
we propose a keyword dictionary that
represents \setKeys{} in either $n \lg \sigma + \Ot{k \lg n}$ or $|T| \lg \sigma + \Ot{k w}$ bits of space, where
   $|T|$ is the number of nodes of a trie representing~$\setKeys{}$.
It supports all operations in $\Oh{m / \alpha + \lg \alpha}$ expected time on an input string of length~$m$ in the word RAM model.
A practical evaluation highlights the practical usefulness of the proposed data structure,
especially for prefix searches --- one of the most essential keyword dictionary operations.
\end{abstract}


\section{Introduction}
\label{sec:introduction}

A \teigi{keyword}~$K$ is a string that is uniquely associated with an integer called the \teigi{identifier} of~$K$.
A \teigi{keyword dictionary} is a data structure that maintains a dynamic set of keywords \setKeys{},
and provides the following operations for a string~$S$ on it:
\begin{itemize}
	\item $\fnInsert(S,i)$ inserts~$S$ into \setKeys{} and assigns~$S$ the identifier~$i$.
	  The identifier is a constant integer unique among all keywords stored in \setKeys{}.
	\item $\fnLookup(S)$ returns the identifier of~$S$ if $S \in \setKeys{}$, or returns the invalid identifier~$\bot$ otherwise.
	\item $\fnDelete(K)$ removes the keyword $K$ from $\setKeys$.
	\item $\fnLocatePrefix(S)$ returns an iterator on the set of identifiers of all keyword in $\setKeys$ having $S$ as a prefix. The iterator can report the next occurrence in constant time.\footnote{We return an iterator instead of this set, since most of the later explained data structures support all operations in the same time $\Oh{t}$ for some $t$, while this operation would take \Oh{t + s} time, if the returned set has size~$s$.}
\end{itemize}

Unlike standard string dictionaries, we omit the operation $\fnAccess{}(i)$ returning the keyword of an identifier~$i$,
as this function can be realized by a separate data structure
(in case of a trie, e.g., an array of pointers in which the $i$-th entry points to the node of the trie representing the keyword~$\fnAccess{}(i)$).
For the performance of practical keyword dictionaries like RDF stores~(e.g., \cite{mavlyutov15diplodocus}), insertions, lookups, and prefix queries are the most crucial operations, on which we want to focus in this article.

A space-efficient representation of a keyword-dictionary is the \emph{compact trie}, in which all unary trie paths are merged such that each internal node has at least two children.
In applications such as information retrieval~\cite{hsu13topk} and database systems~\cite{mavlyutov15diplodocus,zhang18surf},
when designing compact tries,
the main focus is put on properties such as its size and the operation time for insertions, lookups, and prefix queries.
In this article, we propose a new compact trie tailored for excelling at these properties.

\subsection{Preliminaries} \label{secPrelims}
Let $\lg$ denote the logarithm to the base two.
Our model of computation is the standard word RAM model of word size $w$.
We can read and process $\Oh{w}$ bits in constant time.
Let $n$ be a natural number with $n = \Oh{2^w}$.
Storing an integer of the domain $[1..n]$ costs $\lg n$ bits
such that pointers for the problem size~$n$ can be represented in $\lg n$ bits (like in the transdichotomous model).
The choice of this model (severing the connection between word size and the logarithm of the problem size) is justified by the fact that the register sizes of SIMD instruction sets is increasing since the recent years significantly (e.g., AVX512 with 512-bit registers).

Let $\Sigma$ be an integer alphabet of size $\sigma \le 2^w$.
An element of $\Sigma^{\ast}$ is called a \teigi{string}.
The length of a string $S$ is denoted by $|S|$.
We write $S[i]$ for the $i$-th character of $S$, for $1 \le i \le |S|$.
The \teigi{empty string} is the string with length zero.
For a string $S = XYZ$, $X$, $Y$, and $Z$ are called a \teigi{prefix}, \teigi{substring}, and \teigi{suffix} of $S$, respectively.
The word RAM model allows us to process $\alpha = \Oh{w / \lg \sigma}$ characters in constant time.

For the rest of the article, let $\setKeys{}$ denote a set consisting of $k$ keywords with a total length of $n = \sum_{K \in \setKeys} |K|$.
We suppose that $\setKeys{}$ is dynamic, and that the integers $k$ and $n$ are variable.
The keywords of~$\setKeys{}$ do not have to be prefix-free.

\begin{table}
	\centerline{\begin{tabular}{llll}
		\hline
		Trie                                    & Space in Bits & Setting\\
		\hline
		\TriePP{}                               & $n   \lg \sigma + {\Oh{k \lg n}}$ & 1 \\
		\TriePP{}                               & $|T| \lg \sigma + {\Oh{k w}}$ & 2 \\
		compact trie                            & $|T| \lg \sigma + {\Oh{k \lg |T|}}$ & 2\\
		z-fast trie~\cite{belazzougui10zfast}   & $|T| \lg \sigma + {\Oh{k w}}$ & 2 \\
		c-packed trie~\cite{takagi17packedtrie} & $|T| \lg \sigma + {\Oh{k w}}$ & 2\\
		\hline
	\end{tabular}
	}\caption{Space complexities of the packed tries addressed in Sect.~\ref{secRelatedWork} under different settings: 
In Setting~1, we concatenate all keywords to a large string of length~$n$.
In this large string, we can address every substring with two pointers of $n$ bits.
We omit Setting~1 for the other compact trie data structures as these (except the plain compact trie) use auxiliary data structures taking ${\protect\Oh{k w}}$ bits.
In Setting~2, we represent each keyword~$K$ with front coding~{\protect\cite[Sect.~4.1]{witten99managing}}, i.e., we represent~$K$ by~{$K[\ell+1..|K|]$} if the longest common prefix of~$K$ with its lexicographically preceding keyword in~$\protect\setKeys$ is $\ell$.
Hence, we store the suffix~{$K[\ell+1..|K|]$} in the trie data structure explicitly as a string. 
By doing so for each keyword, we store $k$ strings with a total length of $|T|$.
Except the plain compact trie, all listed tries store additionally parts of a keyword in {$\protect\Oh{w}$} bits (to apply word-packing techniques),
causing {$\protect\Oh{kw}$} bits of additional space.
	}
	\label{tableSpaceComplexity}
\end{table}

\subsection{Related Work}\label{secRelatedWork}
Keyword dictionaries are an integral data structure with a plethora of applications
(e.g., $n$-gram language models~\cite{pibiri17ngram},
compression~\cite{fischer17lz78},
input method editors~\cite{kudo11language},
query auto-completion~\cite{hsu13topk}, or
range query filtering~\cite{zhang18surf}).
As a well-studied abstract data type they also have many representations.
We refer to standard literature like \cite[Chapter 5.2]{sedgewick14algorithms}, \cite[Chapter 28]{mehta04handbook}, or \cite[Chapter 8.5.3]{navarro16compact} for an introduction to common representations like tries.
Here, we highlight some of the most recent representations, which are all (compact) tries.
A major design choice for a trie is whether to \emph{compact} the unary edges,
spawning two lines of research, which we want to analyze in the following.

For the analysis, let $|T| \le n$ denote the number of nodes of a trie~$T$ storing~$\setKeys$, and
let $m$ be the length of an input string for one of the trie operations.
We start with the non-compact representations:
\begin{itemize}
	\item The \teigi{HAT-trie}~\cite{askitis10trie} is a practically optimized version of the burst trie~\cite{heinz02burst}.
		It suppresses the number of trie nodes by selectively collapsing subtries into cache-conscious hash tables of strings \cite{askitis05arrayhash}.
		Although there is no discussion of prefix searches in \cite{askitis10trie},
		the implementation of Tessil\footnote{\url{https://github.com/Tessil/hat-trie}} supports \fnLocatePrefix{}.
		We are unaware of any theoretical results regarding space or time.

	\item The double array \cite{aoe89doublearray} simulates a trie by using two integer arrays to find a child in constant time,
		and thus can perform \fnLookup{} in $\Oh{m}$ time.
		Although the double array includes some vacant slots and consumes $\Om{n \lg n}$ bits, those vacant slots have a negligible memory effect in practical implementations such as the Cedar trie~\cite{yoshinaga14cedar}.
		In the static setting, Kanda et al.~\cite{kanda17compressed} proposed a practically compressed data structure for the two arrays.
		However, for any of these data structures, it is not clear to us what time is needed for answering \fnLocatePrefix{}.

	\item The Bonsai trie \cite{darragh93bonsai} is a trie whose nodes are maintained in a compact hash table~\cite{cleary84cht}.
		Modern variants~\cite{poyias15bonsai} use $\Oh{n \lg \sigma}$ bits of space in expectancy, and perform $\fnInsert{}$ and $\fnLookup{}$ in \Oh{m} expected time. 
		However, it is not clear how to perform \fnLocatePrefix{} efficiently.

	\item Kanda et al.~\cite{kanda17trie} proposed a dynamic variant of the path decomposed trie of Grossi and Ottaviano~\cite{grossi14pdt} by means of \emph{incremental} path decomposition. This dynamic trie supports \fnInsert{} and \fnLookup{} in $\Oh{m}$ expected time.
		However, there is no discussion about prefix searches. Actually, as Kanda et al.'s trie is based on the Bonsai trie, it faces the same problem for \fnLocatePrefix{}.
\end{itemize}

Considering compact tries, we are aware of the following representations:

\begin{itemize}
	\item
		Jansson et at.~\cite{jansson15lz78} presented a dynamic trie using \Oh{|T| \lg \sigma} bits, in which a leaf can be inserted or deleted in \Oh{(\lg \lg |T|)^2/\lg\lg\lg |T|} time.
		This trie can compute a prefix search in \Oh{(m / \log_\sigma |T|)(\lg\lg |T|)^2/\lg\lg\lg |T|} time~\cite[Thm.~1]{jansson15lz78}.
		In an alternative representation, this trie supports insertions and deletions of leaves in
		\Oh{\lg \lg |T|} expected amortized time while supporting a prefix search in \Oh{m / \lg_\sigma |T| + \lg \lg |T|} worst-case time~\cite[Thm.~2]{jansson15lz78}.

	\item The \teigi{(dynamic) z-fast trie} is a keyword dictionary of Belazzougui et al.~\cite{belazzougui10zfast},
		which uses $|T| \lg \sigma + \Ot{k w}$ bits of space,
		and supports all operations in $\Oh{m/\alpha + \lg m + \lg \lg \sigma}$ expected time\footnote{This time bound can be achieved by omitting the jump pointers in~\cite[Sect.~3.4]{belazzougui10zfast} since their maintenance needs additional time. The jump pointers are used to enable additional operations on the trie such as predecessor queries, which we omit in this article.}.
	\item Takagi et al.~\cite{takagi17packedtrie} proposed the \teigi{dynamic packed compact trie},
		whose name we abbreviate to \teigi{packed c-trie}.
		The packed c-trie uses $|T| \lg \sigma + \Ot{k w}$ bits of space,
		and supports all operations in $\Oh{m/\alpha + \lg w}$ expected time.
	\item HOT~\cite{binna18hot} is an algorithmically engineered trie that applied different strategies
		depending on the distribution of the common prefix lengths of the keywords to obtain high fanouts and
		minimize the depth of the trie.
		It also applies AVX2 instructions for lookup queries.
\end{itemize}

The following keyword dictionaries are static, but share common traits with our proposed data structure:

\begin{itemize}
	\item Grossi and Ottaviano \cite{grossi14pdt} proposed a cache-friendly trie dictionary through path decomposition \cite{ferragina08searching}.
		An operation can be carried out in $\Oh{m + h \log \sigma}$ time, where $h$ is the height of the path-decomposed trie. The data structure is stored in compressed space by exploiting text compression techniques and succinct data structures.
	\item The Marisa trie, developed by Yata \cite{yata2011prefix}, is a static trie that consists of recursively compressed Patricia tries stored in the level-order unary degree sequences (LOUDS) representation~\cite{jacobson89rank}.
		It recursively encodes edge labels in a Patricia trie using another Patricia trie.
		Yata's implementation\footnote{\url{https://github.com/s-yata/marisa-trie}} supports prefix searches.
	\item Arz and Fischer \cite{arz18lzstringdic} proposed a static compressed trie by adapting the LZ78 trie 
		to basic dictionary operations such as \fnLookup{}.
		Their trie uses $\Oh{k \lg n + n \lg \sigma}$ bits of space.
		It can answer \fnLookup{} in \Oh{m} expected time.
		However, we are not aware of whether this data structure supports efficient prefix searches.
	\item Bille et al.~\cite{bille17trie} presented a static keyword dictionary using \Oh{n \lg n} bits of space and \Oh{n} time to represent~$\setKeys{}$.
		It supports queries in \Oh{m/\alpha + \lg m + \lg \lg \sigma} time.
	\item A recent approach is due to  Bille et al.~\cite{bille19top}, who proposed a static keyword dictionary with $\Oh{n \lg \sigma}$ bits of space
		using $\Oh{\min(m \lg \sigma, m + \lg n)}$ time for an operation in the pointer machine model.
	\item The fast succinct trie (FST) is a trie data structure used in the succinct range filter~\cite{zhang18surf}. An FST is divided into two layers at a specific height. The top layer is represented by a \emph{speed}-optimized trie while the bottom layer is represented by a \emph{space}-optimized trie.
		Both tries are represented in the LOUDS representation~\cite{jacobson89rank}.
\end{itemize}

In this article, we present a new keyword dictionary based on the compact trie:

\begin{theorem}\label{thmTrieResult}
	Given a dynamic set $\setKeys{}$ of $k$ keywords whose characters are drawn from an integer alphabet of size $\sigma \le 2^w$,
	there is a keyword dictionary representing~$\setKeys$ in
	either $n \lg \sigma + \Ot{k \lg n}$ or $|T| \lg \sigma + \Ot{k w}$ bits of space, where
	$n  = \sum_{K \in \setKeys} |K|$ is the total length of all keywords of $\setKeys{}$ and
	$|T|$ is the number of nodes of a trie representing~$\setKeys{}$.
	It supports all keyword dictionary operations in $\Oh{m/\alpha + \lg \alpha}$ expected time
	with $\alpha = w/\lg \sigma$ on an input string of length~$m$.
\end{theorem}
Depending on how we represent the input keywords, we obtain two different space complexities in Thm.~\ref{thmTrieResult}, which we put into comparison in Table~\ref{tableSpaceComplexity}.
The time and space bounds of Theorem~\ref{thmTrieResult} are an improvement to all previous compact trie representations we are aware of (e.g.,  the z-fast trie becomes inferior for string lengths $m > w/\lg^2 \sigma$).

Prefix searches arise in various uses of suffix trees,
e.g., computing matching statistics~\cite{gusfield97algorithms},
online suffix tree construction~\cite{Ukkonen95},
online Lempel-Ziv 77 factorization~\cite{LZ77}, just to name a few.
Hence, the time bound for prefix search is of significant theoretical interest,
and our compact trie moves the best known upper bound closer to
the trivial lower bound \Om{m/\alpha}
for reading a pattern of length~$m$ word-packed.
Also, with delete and insert operations,
one can efficiently maintain the \emph{sparse suffix tree}~\cite{karkkainen96sparsest}
for a dynamic set of suffixes to index.

Our experiments in Sect.~\ref{secExperiments} reveal that the above improvements are also practically significant.
We note that other previous trie data structures mentioned earlier have the following drawbacks:
(1) For the HAT-trie or the double array, there are no known nontrivial space and construction time bounds
  as their constructions are based on heuristics. In practice, they are also not favorable for prefix queries.
(2) Trie data structures based on the Bonsai trie have the major drawback that enumerating children is done by querying for each possible edge label in a brute force manner. So they are no-good candidates for prefix search queries, and are therefore omitted in our practical evaluation.
(3) The trie data structure of Jansson et at.~\cite{jansson15lz78} is theoretically appealing, but uses theoretically sophisticated data structures
  for which an efficient implementation seems cumbersome.

\section{Keyword Dictionary \protect\TriePP{}}

\begin{figure}[tb]
	\centering
	\includegraphics[width=10cm,pagebox=cropbox,clip]{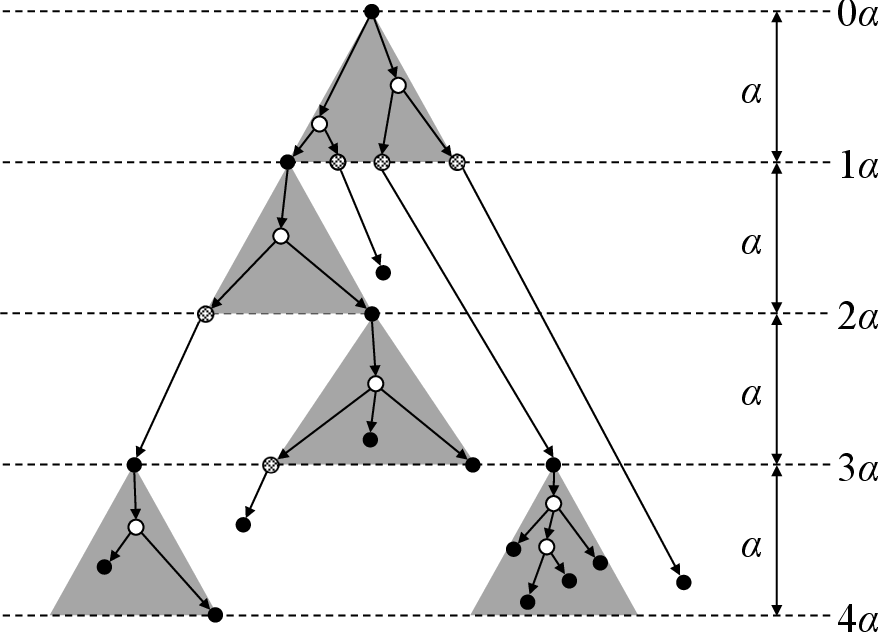}
	\caption{The macro trie of a \protect\TriePP{} instance.
		Micro tries are represented by shaded triangles (cf.~\cite[Fig.~2]{takagi17packedtrie}).
		Circles filled with black color are macro trie nodes.
		Hollow circles are nodes stored exclusively in a micro trie.
		Cross-hatched circles are nodes of a micro trie that are not present in the standard compact trie (as they have only one child).
		These nodes are leaves of a micro trie, and are needed for navigating between the micro trie and the macro trie nodes below of it.
	}
	\label{figMacroTrie}
\end{figure}

\begin{figure}[tb]
	\centering{\begin{adjustbox}{valign=c}\includegraphics[width=5cm,pagebox=cropbox,clip]{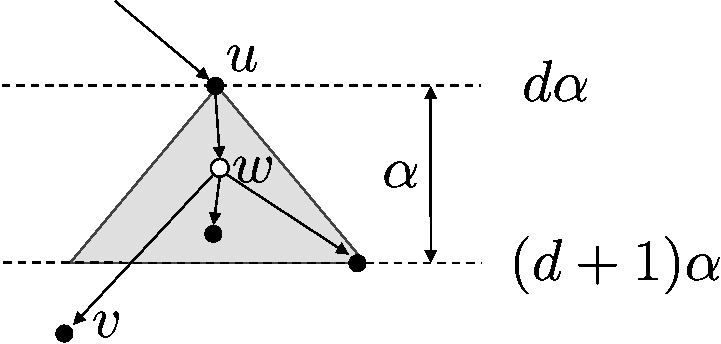}
        \end{adjustbox}
	\hfill
		\begin{adjustbox}{valign=c}$\rightarrow$
        \end{adjustbox}
	\hfill
		\begin{adjustbox}{valign=c}\includegraphics[width=5cm,pagebox=cropbox,clip]{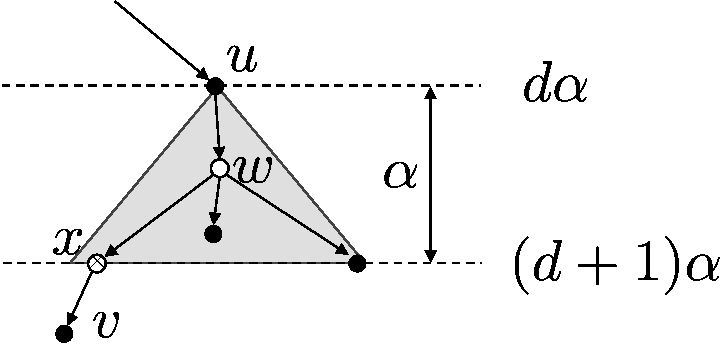}
        \end{adjustbox}
	}\caption{Splitting the edge~$(w,v)$ protruding the boundary of the micro trie rooted at node~$u$ by introducing an artificial micro trie node~$x$. 
	The string depth of $u$ is $d\alpha$ while $v$ with a string depth larger than $(d+1)\alpha$ does not belong to $u$'s micro trie.}
	\label{figAdditionalNodes}
\end{figure}

\begin{figure}[tb]
	\centering
	\includegraphics[width=5cm,pagebox=cropbox,clip]{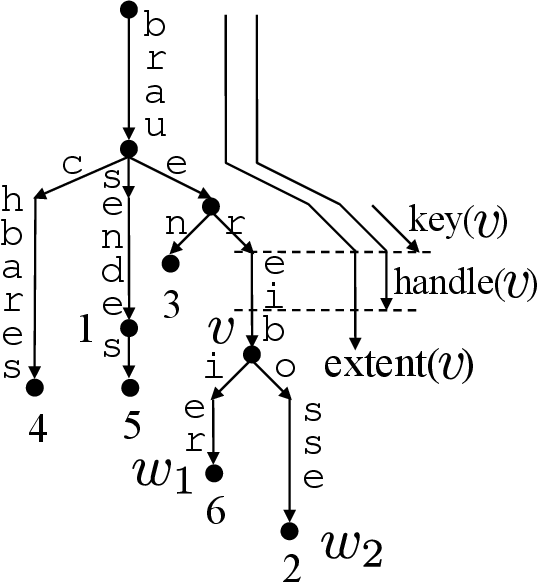}
	\caption{The micro trie built on our running example
		$\setKeys = \{
			K_1 = \texttt{brausende}$,
			$K_2 = \texttt{brauereibosse}$,
			$K_3 = \texttt{brauen}$,
			$K_4 = \texttt{brauchbares}$,
			$K_5 = \texttt{brausendes}$,
			$K_6 = \texttt{brauereibier}
		\}$, which is not prefix-free.
		A leaf~$\nodeU$ storing number~$i$ is associated with the identifier~$i$, i.e., $\protect\fnExtent(\nodeU) = K_i$.
		In this example, the node~\nodeV{} storing the extent \texttt{brauereib} has the leaves~$w_1$ and~$w_2$ representing the keywords~$K_6$ and $K_2$, respectively, as its children,
		which are determined by their keys $\protect\fnKey(w_1) = \texttt{i}$ and $\protect\fnKey(w_2) = \texttt{o}$, respectively.
		If we assume that eight characters fit into a computer word,
		then the extent of~$v$ is outside of the micro trie containing the root node. This fact is symbolized by the dashed line
		separating the eighth and the ninth character of \protect\fnExtent(v).
	}
	\label{figMicroTrie}
\end{figure}

Tuning for fast prefix searches, our idea is to devise a new keyword dictionary based on the compact trie data structures,
as they are practically faster than approaches based on the double array when the prefixes in question are relatively short to the stored keywords.
Our approach, called \TriePP{} for \teigi{improved compact trie}, is a hybrid of the z-fast trie and the packed c-trie.
Like these two trie representations, the compact trie is decomposed in a macro trie storing micro tries.

For a formal explanation of this decomposition, let the \teigi{string depth} of a node~$\nodeU$ denote the length of the concatenation of all labels on the path from the macro trie root to~$\nodeU$.
To keep the following explanation simple, let us assume, for the time being, that the keyword set~$\setKeys$ is prefix-free such that each leaf corresponds to one keyword. (In the general case, we do not only consider leaves but also internal nodes corresponding to a keyword.)
Our starting point is a compact trie.
If there is an edge leading to an internal node, we split up this edge by creating additional nodes on this edge whose string depths are multiples of $\alpha$.
Subsequently, we put all nodes whose string depths are multiples of $\alpha$ into the macro trie.
Let $\nodeU$ be one of these nodes, and let $d\alpha$ be its string depth.
Then $\nodeU$ becomes the root of a micro trie if it has more than one descendant in the compact trie
whose string depth is at most $(d+1)\alpha$.
Suppose that $\nodeU$ is the root of a micro trie, then this micro trie stores all of $\nodeU$'s descendants (of the compact trie)  whose string depths are at most $(d+1)\alpha$.
Every edge~$(\nodeW,\nodeV)$ from a node~$\nodeW$ of $\nodeU$'s micro trie leading to a descendant~$\nodeV$ of $\nodeU$ with a string depth larger than $(d+1)\alpha$ is split to $(\nodeW,x)$ and $(x,\nodeV)$ for an artificial node~$x$ with string depth $(d+1)\alpha$ (cf.\ the cross-hatched circles in Fig.~\ref{figMacroTrie} and Fig.~\ref{figAdditionalNodes} for a schematic illustration).
Finally, leaves of the compact trie are macro trie nodes.
As previously explained, there can additionally be micro trie nodes if (a) their string depths are between $d\alpha$ and $(d+1)\alpha$ and (b) they have an ancestor with string depth~$d\alpha$ that is the root of the respective micro trie.
Consequently, the total number of micro and macro trie nodes is bounded by \Oh{k}, where $k$ is the number of nodes in the compact trie.
Fig.~\ref{figMacroTrie} captures this schematically.

For \TriePP{}, we apply the explained trie decomposition, which coincides with the trie decomposition of the packed c-trie for the macro trie.
Our micro tries are \emph{alphabet-aware z-fast tries}, whose definition follows.
The z-fast trie proposed by Belazzougui et al.~\cite{belazzougui10zfast} works on binary strings.
Their results on micro trees work for binary strings up to length \Oh{w}.
Here, we propose a variant, the \teigi{alphabet-aware z-fast tries}, 
that manages strings on the alphabet~$\Sigma$ up to length $\Oh{w/\lg \sigma} = \Oh{\alpha}$ by packing $\Oh{\alpha}$ characters in a constant number of machine words:

\begin{lemma}\label{lemMicroTrie}
	Let $\setKeys{}$ be a dynamic set of $k$ keywords whose characters are drawn from an alphabet of size $\sigma \le 2^w$.
	Given that each keyword of $\setKeys{}$ has a length of \Oh{\alpha},
	there is a keyword dictionary representing $\setKeys$ in
	either $n \lg \sigma + \Ot{k \lg n}$ or $|T| \lg \sigma + \Ot{k w}$ bits of space,
	where $\alpha = w/\lg \sigma$,
	$n  = \sum_{K \in \setKeys} |K| \le \alpha |\setKeys|$ is the total length of all keywords of $\setKeys{}$, and
	$|T|$ is the number of nodes of a trie representing~$\setKeys{}$.
	It supports all keyword dictionary operations
	in either \Oh{\lg \alpha} expected time or \Oh{\lg \alpha \lg^2 \lg \sigma / \lg \lg \lg \sigma} deterministic time.
\end{lemma}

An operation with a string of length~$m$ with $m = \Om{\alpha}$ (but with $m = \Oh{2^w}$) involves the traversal of the macro tree,
which is done in \Oh{m/\alpha} expected time\footnote{See Sect.~\ref{secMacro} for a detailed description of the macro trie.}
for all keyword dictionary operations~\cite{takagi17packedtrie}.
Combining the operations in the macro trie and in the micro tries gives \Oh{m / \alpha + \lg \alpha} total time, and concludes Theorem~\ref{thmTrieResult}.

\subsection{Micro Tries}
For explaining \TriePP{} in detail, we start with a review of the z-fast trie under the light of our alphabet-aware variant.
We say that a node~$v$ is \teigi{associated with the identifier} of a keyword~$K$ if we can read~$K$ by following the path from the root to~$v$.
The alphabet-aware z-fast trie is a compact trie in which each leaf~$v$ is associated with the identifier of a keyword.
An internal node has at least two children unless it is also associated with the identifier of a keyword.
If the set of keywords~\setKeys{} is prefix-free, then there are no nodes with a single child.

Fig.~\ref{figMicroTrie} presents an instance of such a trie.
The figure also depicts the following definitions that are substrings or nodes associated to each node of an alphabet-aware z-fast trie.

\begin{figure}
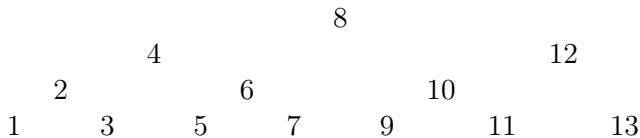

	\centering{\begin{tabular}{*{13}{c}}
			  &  &   &  &   &  &   & 8 &   &  &    &  \\
			  &  &   & 4 &   &  &   &  &   &  &    & 12 \\
			  &2 &   &  &   & 6&   &  &   &10  &    &  \\
			1 &  & 3 &  & 5 &  & 7 &  & 9 &  & 11 &  & 13 \\
		\end{tabular}
	}\caption{Geometric interpretation of the 2-fattest numbers.
	Based on the number line of natural numbers, we move each natural number as many position upwards as it has trailing zeros in its binary representation.
In this example of numbers from 1 to 13, the highest number is 8.
Now the 2-fattest number of a query interval is the highest number within this range.
}
	\label{figFattestNumber}
\end{figure}

\begin{itemize}
	\item $\fnKey(\nodeV)$ is the first character in the label of the edge connecting $\nodeV$ with its parent.
		It is undefined if \nodeV{} is the root.
	\item $\fnExtent(\nodeV)$ is the string obtained by concatenating the edge labels of the path from the root node to $\nodeV$.
	\item $\fnExit(S)$ is the highest node $\nodeV$ for which, among all other nodes,
		the longest common prefix between $S$ and $\fnExtent(\nodeV)$ is the longest.
	\item $\fnParex(S)$ is the parent node of $\fnExit(S)$,
		or a special symbol $\bot$ with $\fnExtent(\bot) = 1$ if $\fnExit(S)$ is the root node.
\end{itemize}
It is left to explain for what $\fnHandle(\nodeV)$ stands in the figure.
For that we need the notion of 2-fattest numbers~\cite[Def.\ 1]{belazzougui10zfast}.
The \teigi{2-fattest number} of an interval $[\ell..r]$ of positive integers $0 < \ell \le r$ is the integer in $[\ell..r]$
with the most trailing zeros in its binary representation (see Fig.~\ref{figFattestNumber} for a geometric interpretation).
Given a node \nodeV{} with its parent \nodeU{}, we can compute
the 2-fattest number~$f$ of $[|\fnExtent(\nodeU)| + 1..|\fnExtent(\nodeV)|]$ to
determine the handle of~\nodeV{}, which is $\fnHandle(\nodeV) := \fnExtent(\nodeV)[1..f]$.
In case that \nodeV{} is the root, we set $\fnHandle(\nodeV)$ to the empty string.

For supporting the keyword dictionary operations, we need operations to descend in a micro tree.
For that, the trie maintains a dictionary \DicHandle{} that can address each internal node~$\nodeU$ by its handle $\fnHandle(\nodeU)$.
Additionally, we need a way to navigate from a node to one of its children.
This can be done in constant time in the original z-fast trie since it works on a binary alphabet (hence, each node has at most two children).
For the alphabet-aware variant, 
each internal node~\nodeU{} stores a dictionary \DicChild{\nodeU} to access one of its child nodes $v$ by the character $\fnKey(\nodeV)$.
The proof of Lemma~\ref{lemMicroTrie} gives us two different, possible representations for \DicChild{}:

\begin{proof}[Proof of Lemma~\ref{lemMicroTrie}]
	Since the edge labels in the alphabet-aware z-fast trie are characters drawn from the integer alphabet~$\Sigma$,
	traversing from a node to a specific child costs \Oh{\sigma} time.
	We improve this time by augmenting each node with a data structure maintaining its children such that, given a node~\nodeV{} and a character~$c$, we can navigate from \nodeV{} to its child connected with the edge starting with~$c$ by querying this data structure
	having stored $c$ and $\nodeV{}$ as key and value, respectively.
    This data structure can be realized with a hash table with constant expected time, or with a dynamic predecessor data structure like~\cite{beame02predecessor} (combined with the transformation of Andersson and Thorup~\cite{andersson00transform}) taking \Oh{m \lg n} bits
	and supporting all operations in
	$\Oh{\lg \lg \sigma \lg \lg m / \lg \lg \lg \sigma} = \Oh{\lg^2 \lg \sigma / \lg \lg \lg \sigma}$ deterministic time when storing $m \le \sigma$ elements (the space bounds are due to the fact that we store pointers to the specific children as satellite data).
	This sums up to \Oh{k \lg n} bits as we have \Oh{k} trie nodes.
\end{proof}

For the algorithmic part, we follow Algorithm~1 and Section~3.3 of \cite{belazzougui10zfast}.
Given a pattern~$P$ of length~$\Oh{\alpha}$, this algorithm locates \fnExit(P) and \fnParex(P).
Having  \fnExit(P) and \fnParex(P), we can perform all keyword dictionary operations as in the z-fast trie.
The idea of the algorithm is to perform a search on the interval $[\ell..r]$, which is set to $[1..|P|]$ at the beginning.
The search handles this interval similarly to a binary search with the aim to find the lowest node whose handle is a prefix of~$P$.
For explanation, the algorithm is divided into rounds.
In each round, it (a) either enlarges~$\ell$ or shrinks~$r$, (b) computes the 2-fattest number~$f$ of $[\ell..r]$,
and (c) queries \DicHandle{} with the handle~$P[1..f]$.
If there is a node~$v$ with $\fnHandle(v) = P[1..f]$,
the algorithm has matched $P[1..f]$ with this node and simulates the descending to this trie node
by setting $\ell \gets |\fnExtent(v)|$.
Otherwise (there is no such node~$v$), the algorithm sets $r \gets f-1$ to aim for jumping to a node whose extent is less than $f$.
The algorithm stops when it finds either \fnExit(P) and \fnParex(P)~\cite[Thm.~3]{belazzougui10zfast}, which is after \Oh{\lg |P|} rounds.
If \fnExit(P) is found, it has previously already computed \fnParex(P).
Otherwise, it takes that child of \fnParex(P) whose edge connected to \fnParex(P) leads us to \fnExit(P).
For finding this child, the algorithm uses \DicChild{\fnParex(P)}.
Finally, the updates can be conducted with a constant number of pointer updates (detailed are described in \cite[Sect.~5]{belazzougui10zfast}).

In the context of the example of Fig.~\ref{figMicroTrie}, this algorithm applied to $P = \texttt{brauereibock}$ gives us the node~\fnExit(P), which is the node~\nodeV{} visualized in Fig.~\ref{figMicroTrie}.
From there, we can query \DicChild{\fnExit(P)} for the predecessor (resp.\ successor) with the character~\texttt{o} to find
the predecessor (resp.\ successor) of $P$, which is $K_6$ (resp.\ $K_2$).

\subsection{Macro Trie}\label{secMacro}
It is left to describe the macro trie borrowed from the packed c-trie, and to analyze
the space and time complexity of \TriePP{}.
The macro trie is needed to cope with keywords longer than $\alpha$~characters, or $w$ bits.
The rough idea is to partition a long keyword into chunks of $w$ bits,
and maintain the chunks in a dictionary \DicChunk{} similar to \DicHandle{}, mapping $w$-bit chunks to macro trie nodes.
Given that the root is at height~$0$,
a node on a height $h$ of the macro trie is endowed with
\begin{itemize}
	\item a micro trie representing its descendants whose extents are at most $(h+1)w$ bits long, and with
	\item a \DicChunk{} representing its children whose extents are longer than $(h+1)w$ bits.
\end{itemize}
Its \DicChunk{} stores the $w$-bit substring starting at the $(hw+1)$-th bit of the extents of its respective children,
where the \teigi{extent of a macro trie node}~$v$ is the binary representation of the string read from the path from the macro tree root to~$v$.
(Consequently, the string depth of a node is the length of its extent.)
An update of the trie involves a lookup of the insertion or deletion position, and a modification of a \DicChunk{} or a micro trie.

\paragraph{Space Complexity}
Our keyword dictionary~\TriePP{} maintains \Oh{k} macro and \Oh{k} micro nodes.
Each node stores a pointer to a substring of a keyword.
The keywords are stored either in a concatenated string of length~$n \lg \sigma$,
or are compressed via front coding~\cite[Sect.~4.1]{witten99managing} taking $|T| \lg \sigma$ bits in total.
We store $\fnExtent(v)$ of a node~$v$ either as two $n$-bit pointers to the concatenated string (former case) or verbatim in $w$-bits (latter case).
Since the number of total nodes stored in the \DicChild{}s, the \DicHandle{}s and the \DicChunk{}s is \Oh{k},
the data structure needs in total either $n \lg \sigma + \Oh{k \lg n}$ or $|T| \lg \sigma + \Ot{k w}$ bits.
This gives the bounds in Table~\ref{tableSpaceComplexity}.

\paragraph{Time Complexity}
Given a pattern~$P$ of length~$m$,
we can traverse the macro trie by visiting at most $m/\alpha$ macro trie nodes to find the micro trie~$\tau$ storing
the node whose extent has the longest common prefix with~$P$.
After reaching $\tau$,
we can compute the handle of a node from its extent in constant time, since the 2-fattest number in $[\ell..r]$ is
the integer $(-1~\texttt{<<}~\texttt{msb}( (\ell-1) \mathbin{\oplus} r )) \mathbin{\&} r$, where \texttt{<<}, \texttt{msb}, $\oplus$ and $\&$ denote the bitwise left shift, the function retrieving the most significant bit, the bitwise exclusive-\texttt{OR} and the bitwise \texttt{AND} operators, respectively.
In total, we query \Oh{m/\alpha} \DicChunk{}s, $\tau$'s \DicHandle{} \Oh{\lg \alpha} times, and \DicChild{\fnParex(P)} at most one time, yielding
\Oh{m/\alpha + \lg \alpha} expected time as claimed in Theorem~\ref{thmTrieResult} if all dictionaries can lookup an entry in constant expected time.
Choosing a suitable representation for \DicHandle{}, \DicChild{}, and \DicChunk{} is the major task of the next subsection dealing
with practical aspects of \TriePP{}.

\subsection{Implementation Techniques}\label{secImplementation}
On the practical side, our major improvements are based on the three ideas:
\begin{enumerate}[label=Task~{\arabic*},ref=Task~{\arabic*}]
	\item representing each node by an identifier (ID) to store IDs instead of node pointers,  \label{itNodeToID}
	\item storing a global mapping from $\fnExtent(\nodeV)$ to node IDs, and \label{itGlobalMapping}
	\item representing the dictionaries with different data structures with focus on either speed or memory efficiency.
		\label{itDifferentDict}
\end{enumerate}

\paragraph{Micro Tries}
Each node \nodeV{} stores~$\fnExtent(\nodeV)$, which can be represented in a constant number of computer words.
From $\fnExtent(\nodeV)$ we can deduce $\fnHandle(\nodeV)$ and $\fnKey(\nodeV)$ in constant time.
Therefore, the dictionaries \DicChild{} and \DicHandle{} have no need to store the keys of their entries
since it suffices to maintain the nodes with which a dictionary can restore the respective keys on demand.
By doing so, a lookup of a node $\nodeV$ with a key $\fnHandle(\nodeV)$ (resp.\ $\fnKey(\nodeV)$) needs to compute
$\fnHandle(\nodeW)$ (resp.\ $\fnKey(\nodeW)$) of each node~\nodeW{} in question for comparison.
By conducting the comparisons in this way, we save memory by omitting the keys at the expense that the benefits of current processors featuring large cache lines become negligible in this context.
(Imagine a linear probing hash table storing the keys explicitly, where we can fetch keys stored at successive cells at once for collision probing.)
Here, we embrace the cuckoo hashing~\cite{pagh04cuckoo} technique, which has strong theoretical results in the pointer machine model.
This concludes our approach for \ref{itNodeToID}.

\paragraph{Node Factory}
In our setting, we assume that $k$ is much small than~$n$.
Otherwise, \TriePP{} becomes unfavorable with respect to other trie data structures like the Bonsai trie.
That is because our trie data structure contains \Ot{k} nodes in total. However, using $w$ bits for a pointer to a node is wasteful.
Instead, we want to store node pointers in $\Ot{\lg k}$ bits as hinted in the description of our computational model in Sect.~\ref{secPrelims}.
For that (as highlighted in \ref{itGlobalMapping}), we store each node in a global two-dimensional array that assigns each node an integer represented in $\Ot{\lg k}$ bits,
which we set to 32 bits for the experiments.
By storing 32-bit integers instead of pointers on commodity computers with a word size of $w = 64$ bits,
we can roughly halve the memory requirement for maintaining \DicChild{} and \DicHandle{}.

\paragraph{Macro Trie}
Like for \DicHandle{}, we use a cuckoo hash table for representing the \DicChunk{}s.
We again just store the nodes in the cuckoo hash table, since we can restore their keys by extracting the respective $w$-bit substring
in constant time.
We also maintain a separate node factory storing the macro trie nodes.

\paragraph{Cuckoo Hashing and Practical Considerations}
Our cuckoo hash table~$H$ uses three hash functions.
We restrict the hash table size~$|H|$ to be a power of two 
such that we can map a hash value to $[1..|H|]$ more quickly by using bit shifts instead of a modulo operation
(cf.\ the discussion in \cite[Sect.\ 1]{sheldon01division}; however, new techniques~\cite{lemire19remainder} can speed this up).
An insertion collision occurs if each of the entries located by the hash functions is already occupied.
Given such a collision on inserting an element~$e$, we start a random walk by selecting the $i$-th hash functions~$h_i$ for a random~$i$, swapping
$H[h_i]$ with $e$ and recurse.
If this walk is unsuccessful after a certain number of steps, the hash table doubles its size.
To keep the memory requirement at minimum, the chosen hash functions are determined at startup and are the same across all cuckoo hash tables.
The hash functions are based on three xorshift operations borrowed from MurmurHash\footnote{\url{https://github.com/aappleby/smhasher/wiki/MurmurHash3}} and two multiplications with different 64-bit integer seeds.
Unwisely chosen seeds can result in a failure of the data structure, as the hash functions are immutable (changing would cause to rehash \emph{all} cuckoo hash table instances).
However, this was not a problem in our experiments.
While insertions take \Oh{1} expected time for a sufficiently small \teigi{load factor}, i.e., the maximum ratio between the number of stored elements and $|H|$ before doubling the size of $H$, a lookup takes \Oh{1} worst case time.
The load factor in combination with the threshold on the maximal number of iterations for collision handling
does not have much influence on the final size, since a higher load factor
makes it more probable that an insertion collision exceeds the threshold.
Setting this threshold to a smaller value boosts the insertion speed at the expense of a higher risk of creating an unnecessarily large table.
However, preliminary experiments were in favor for a small threshold around 100 iterations.
For the experiments in the following section, we fixed the theshold to 100, and set the load factor to $0.9$.

\paragraph{First-Child Next-Sibling Representation}
In practice, the Cuckoo hash tables used for representing the dictionaries \DicChild{} waste non-negligible space as
(a) each micro trie node stores such a hash table, and (b) the hash tables may not always become full.
For space efficiency, we did not follow this approach, but instead represent all \DicChild{} dictionaries of a micro trie
with a single trie data structure in the \teigi{first-child next-sibling (FNCS)} representation (see~\cite{lovrencic08fcns} for a definition).
In this representation, we maintain two arrays for (a) the first children and (b) the next siblings,
where (a) and (b) are pointers gained from the node factory.
For navigation in the FNCS representation it is necessary to know the character of the in-going edge of each node~$\nodeV$,
but this information is already given by querying $\fnKey(\nodeV)$.
This concludes our strategy for \ref{itDifferentDict}.

\paragraph{Number of Nodes}
The implementations of the compact trie, the packed c-trie, the z-fast trie, and \protect\TriePP{} have same number of nodes.
This seems to contradict the prior statement (cf.\ Fig.~\ref{figAdditionalNodes}) that the packed c-trie and \protect\TriePP{} additionally introduce nodes at string depths $k\alpha$ for an integer~$k$.
However, we introduced these nodes only for didactic reasons.
The actual implementations do not create these nodes as we can use the respective (actually created) descendants of these nodes as well.

\begin{table}\centering
	\scriptsize
    \begin{tabular}{|l*{9}{|r}|}
		\hline
		\multicolumn{1}{|c}{\setKeys{}}&
		\multicolumn{1}{|c}{$\frac{n}{10^6}$} &
		\multicolumn{1}{|c}{$\sigma$} &
		\multicolumn{1}{|c}{$\frac{k}{10^3}$} &
		\multicolumn{1}{|c}{$\varnothing$len} &
		\multicolumn{1}{|c}{max-len} &
		\multicolumn{1}{|c}{$\varnothing$LCP} &
		\multicolumn{1}{|c}{max-LCP} &
		\multicolumn{1}{|c}{$\frac{|T|}{10^6}$} &
		\multicolumn{1}{|c|}{$\frac{|C|}{10^3}$}  \\
		\hline
		\texttt{proteins   } &   903 &  26 &  2,982 & 302.8 &    36,805 & 38.8 &  16,190 & 787 &  5,778 \\
		\texttt{urls       } & 1,413 &  98 & 18,564 &  76.1 &     2,048 & 60.9 &   2,006 & 282 & 35,343 \\
		\texttt{dblp.xml   } &   169 &  96 &  2,950 &  57.6 &       685 & 34.4 &     104 &  68 &  5,900 \\
		\texttt{geographic } &   107 & 134 &  7,308 &  14.6 &       151 &  8.5 &     247 &  45 & 12,802 \\
		\texttt{commoncrawl} &   121 & 113 &  1,995 &  61.0 & 1,194,988 & 12.9 & 119,276 &  96 &  3,740 \\
		\texttt{vital      } &   243 & 203 &    494 & 493.3 &     9,794 & 12.7 &   1,806 & 238 &    986 \\
		\hline
	\end{tabular}
	\caption{Characteristics of our keyword sets. The total length of all keywords is $n$.
		The number of keywords is $k$.
		The average and maximum length of a keyword is written in the columns \emph{$\varnothing$len} and \emph{max-len}, respectively.
		The columns \emph{$\varnothing$LCP} and \emph{max-LCP} show, respectively, the average length and the maximal length of the longest common prefixes of all keywords.
		The number of nodes a compact trie~$C$ stores is given by~$|C|$. 
	}
	\label{tabDatasets}
\end{table}

\begin{table}
    \scriptsize
	\centering
	\subfloat[$\#len \leftrightarrow |len|$  Histogram]{\label{tabDistributionLen}
		\begin{tabular}{*{9}{r}}
			\hline
			$i$ &
			\multicolumn{1}{c}{\texttt{proteins}} &
			\multicolumn{1}{c}{\texttt{urls}} &
			\multicolumn{1}{c}{\texttt{dblp.xml}} &
			\multicolumn{1}{c}{\texttt{geographic}}  &
			\multicolumn{1}{c}{\texttt{commoncrawl}}  &
			\multicolumn{1}{c}{\texttt{vital}} \\
			\hline
			1 & 19 & 85 & 2 & 11 & 97 & 39 \\
			2 & 132 & 851 & 1 & 262 & 1,546 & 26 \\
			4 & 5,485 & 7,888 & 0 & 31,036 & 31,931 & 131 \\
			8 & 36,973 & 25,921 & 5 & 1,270,765 & 137,074 & 726 \\
			16 & 75,796 & 24,188 & 25 & 3,899,303 & 636,922 & 2,298 \\
			32 & 66,530 & 197,634 & 395,244 & 1,838,186 & 445,153 & 4,932 \\
			64 & 130,527 & 8,620,706 & 1,801,952 & 263,086 & 369,674 & 12,007 \\
			128 & 481,117 & 8,463,502 & 723,011 & 5,398 & 255,830 & 32,038 \\
			256 & 818,538 & 1,100,909 & 29,782 & 7 & 61,018 & 75,871 \\
			512 & 955,403 & 100,867 & 213 & 0 & 36,936 & 166,775 \\
			1,024 & 343,983 & 19,207 & 2 & 0 & 11,627 & 165,169 \\
			2,048 & 57,653 & 2,946 & 0 & 0 & 4,464 & 33,599 \\
			4,096 & 8,691 & 0 & 0 & 0 & 1,878 & 857 \\
			8,192 & 1,145 & 0 & 0 & 0 & 795 & 14 \\
			16,384 & 83 & 0 & 0 & 0 & 256 & 1 \\
			32,768 & 15 & 0 & 0 & 0 & 99 & 0 \\
			65,536 & 2 & 0 & 0 & 0 & 52 & 0 \\
			131,072 & 0 & 0 & 0 & 0 & 39 & 0 \\
			262,144 & 0 & 0 & 0 & 0 & 5 & 0 \\
			524,288 & 0 & 0 & 0 & 0 & 3 & 0 \\
			1,048,576 & 0 & 0 & 0 & 0 & 2 & 0 \\
			2,097,152 & 0 & 0 & 0 & 0 & 1 & 0 \\
			\hline
		\end{tabular}
	}

	\subfloat[$\#LCP \leftrightarrow |LCP|$  Histogram]{\label{tabDistributionLCP}
		\begin{tabular}{*{9}{r}}
			\hline
			$i$ &
			\multicolumn{1}{c}{\texttt{proteins}} &
			\multicolumn{1}{c}{\texttt{urls}} &
			\multicolumn{1}{c}{\texttt{dblp.xml}} &
			\multicolumn{1}{c}{\texttt{geographic}}  &
			\multicolumn{1}{c}{\texttt{commoncrawl}}  &
			\multicolumn{1}{c}{\texttt{vital}} \\
			\hline
			0 & 22 & 91 & 2 & 84 & 101 & 111 \\
			1 & 490 & 2,633 & 19 & 2,225 & 6,012 & 1,850 \\
			2 & 9,014 & 11,115 & 20 & 19,636 & 50,615 & 6,079 \\
			4 & 470,608 & 29,492 & 5 & 635,924 & 306,121 & 28,013 \\
			8 & 1,432,010 & 26,723 & 2,663 & 3,838,361 & 574,787 & 118,627 \\
			16 & 203,019 & 76,180 & 556,906 & 2,457,041 & 780,370 & 240,884 \\
			32 & 207,474 & 1,459,143 & 862,179 & 319,203 & 173,276 & 92,179 \\
			64 & 205,067 & 10,668,966 & 1,398,593 & 34,830 & 77,137 & 4,730 \\
			128 & 204,307 & 5,814,357 & 129,715 & 749 & 21,026 & 1,043 \\
			256 & 155,849 & 429,835 & 134 & 0 & 3,870 & 559 \\
			512 & 73,927 & 37,058 & 0 & 0 & 1,247 & 309 \\
			1,024 & 17,440 & 8,263 & 0 & 0 & 507 & 93 \\
			2,048 & 2,468 & 847 & 0 & 0 & 193 & 5 \\
			4,096 & 335 & 0 & 0 & 0 & 48 & 0 \\
			8,192 & 60 & 0 & 0 & 0 & 18 & 0 \\
			16,384 & 1 & 0 & 0 & 0 & 70 & 0 \\
			32,768 & 0 & 0 & 0 & 0 & 0 & 0 \\
			65,536 & 0 & 0 & 0 & 0 & 0 & 0 \\
			131,072 & 0 & 0 & 0 & 0 & 3 & 0 \\
			\hline
		\end{tabular}
	}\caption{Histogram of (a) keyword lengths and (b) the lengths of the longest common prefixes (LCPs) of the keywords.
		While Table~\ref{tabDatasets} captures the average and maximal lengths of the keywords and their LCPs,
		these tables give an insight in the distributions of the lengths and the LCPs.
		A length is counted in the $i$-th row if is $i$ for $i=1$ and $i=2$, or belongs in $[2^{i-2}+1..2^{i-1}]$ for $i \ge 3$.
	}
	\label{tabDistributionLengths}
\end{table}

\begin{table}
    \scriptsize
    \centering
    \subfloat[$\#\DicHandle{} \leftrightarrow |\DicHandle{}|$  Histogram]{\label{tabDistributionHandle}
		\begin{tabular}{*{9}{r}}
			\hline
            $i$ &
			\multicolumn{1}{c}{\texttt{proteins}} &
			\multicolumn{1}{c}{\texttt{urls}} &
			\multicolumn{1}{c}{\texttt{dblp.xml}} &
			\multicolumn{1}{c}{\texttt{geographic}}  &
			\multicolumn{1}{c}{\texttt{commoncrawl}}  &
			\multicolumn{1}{c}{\texttt{vital}} \\
            \hline
			1 & 692,786 & 1,996,651 & 233,983 & 474,823 & 180107 & 57649 \\
			2 & 72,926 & 419,911 & 46,975 & 126,163 & 36273 & 13791 \\
			4 & 26,863 & 278,813 & 27,291 & 70,145 & 19255 & 8641 \\
			8 & 7,696 & 143,852 & 16,392 & 30,265 & 8500 & 4097 \\
			16 & 1,705 & 66,594 & 1,1386 & 13,357 & 3449 & 1651 \\
			32 & 420 & 27,161 & 6,411 & 6,424 & 1195 & 618 \\
			64 & 89 & 11,108 & 3,214 & 2,952 & 488 & 254 \\
			128 & 24 & 4,574 & 1,152 & 1,241 & 194 & 105 \\
			256 & 5 & 1,633 & 302 & 472 & 100 & 25 \\
			512 & 1 & 580 & 110 & 191 & 38 & 13 \\
			1024 & 1 & 75 & 37 & 68 & 37 & 3 \\
			2048 & 0 & 0 & 18 & 21 & 1 & 0 \\
			4096 & 0 & 0 & 8 & 9 & 0 & 0 \\
			8192 & 0 & 0 & 4 & 0 & 0 & 0 \\
			16384 & 0 & 1 & 0 & 1 & 0 & 0 \\
			32768 & 0 & 1 & 0 & 0 & 0 & 0 \\
			65536 & 0 & 0 & 0 & 0 & 0 & 1 \\
			131072 & 0 & 0 & 1 & 0 & 0 & 0 \\
			262144 & 0 & 0 & 0 & 0 & 0 & 0 \\
			524288 & 0 & 0 & 0 & 0 & 1 & 0 \\
			1048576 & 1 & 0 & 0 & 0 & 0 & 0 \\
			\hline
		\end{tabular}
	}

    \subfloat[$\#\DicChild{} \leftrightarrow |\DicChild{}|$  Histogram]{\label{tabDistributionChild}
		\begin{tabular}{*{9}{r}}
			\hline
            $i$ &
			\multicolumn{1}{c}{\texttt{proteins}} &
			\multicolumn{1}{c}{\texttt{urls}} &
			\multicolumn{1}{c}{\texttt{dblp.xml}} &
			\multicolumn{1}{c}{\texttt{geographic}}  &
			\multicolumn{1}{c}{\texttt{commoncrawl}}  &
			\multicolumn{1}{c}{\texttt{vital}} \\
            \hline
			1 & 27,933 & 189,554 & 106 & 204,565 & 30,276 & 279 \\
			2 & 1,220,896 & 3,939,539 & 808,559 & 1,644,531 & 468,330 & 164,154 \\
			4 & 231,439 & 1,594,225 & 313,011 & 716,500 & 175,809 & 54,646 \\
			8 & 86,483 & 886,825 & 116,020 & 288,994 & 69,654 & 19,579 \\
			16 & 42,571 & 507,437 & 53,258 & 104,526 & 47,619 & 6,272 \\
			32 & 13,894 & 34,609 & 14,298 & 28,445 & 6365 & 1,466 \\
			64 & 0 & 1,221 & 656 & 301 & 1,201 & 283 \\
			128 & 0 & 5 & 7 & 8 & 124 & 7 \\
			\hline
		\end{tabular}
	}
	\caption{Histogram of (a) micro tries or (b) internal micro trie nodes storing a specific number of (a) child nodes or (b) internal nodes representing the sizes of (a) all \DicHandle{} instances or (b) all \DicChild{} instances.
		A (a) micro trie or (b) internal node is counted in the $i$-th row if the number of its stored nodes is $i$ for $i=1$ and $i=2$,
		or in $[2^{i-2}+1..2^{i-1}]$ for $i \ge 3$.
		None of the keyword sets is prefix-free, as can be seen by the fact that there are nodes with only a single child.
	}
	\label{tabHistogramDict}
\end{table}

\section{Experiments}\label{secExperiments}
Finally, we analyze the empirical performance of \TriePP{} with respect to time and memory consumption.
In particular, we are interested in the running time of $\fnInsert{}$, $\fnLookup{}$, $\fnLocatePrefix{}$, and $\fnDelete{}$.
For that, we implemented \TriePP{} in \CPlusPlus{}.
Our implementation of \TriePP{} is written in \CPlusPlus{}, and available at \url{https://gitlab.com/habatakitai/ctriepp}.
For the experiments, we set up a machine equipped with CentOS 6.10, with an Intel Xeon X5560 processor running at 2.80 GHz, and with 198GB of main memory.

\begin{figure}
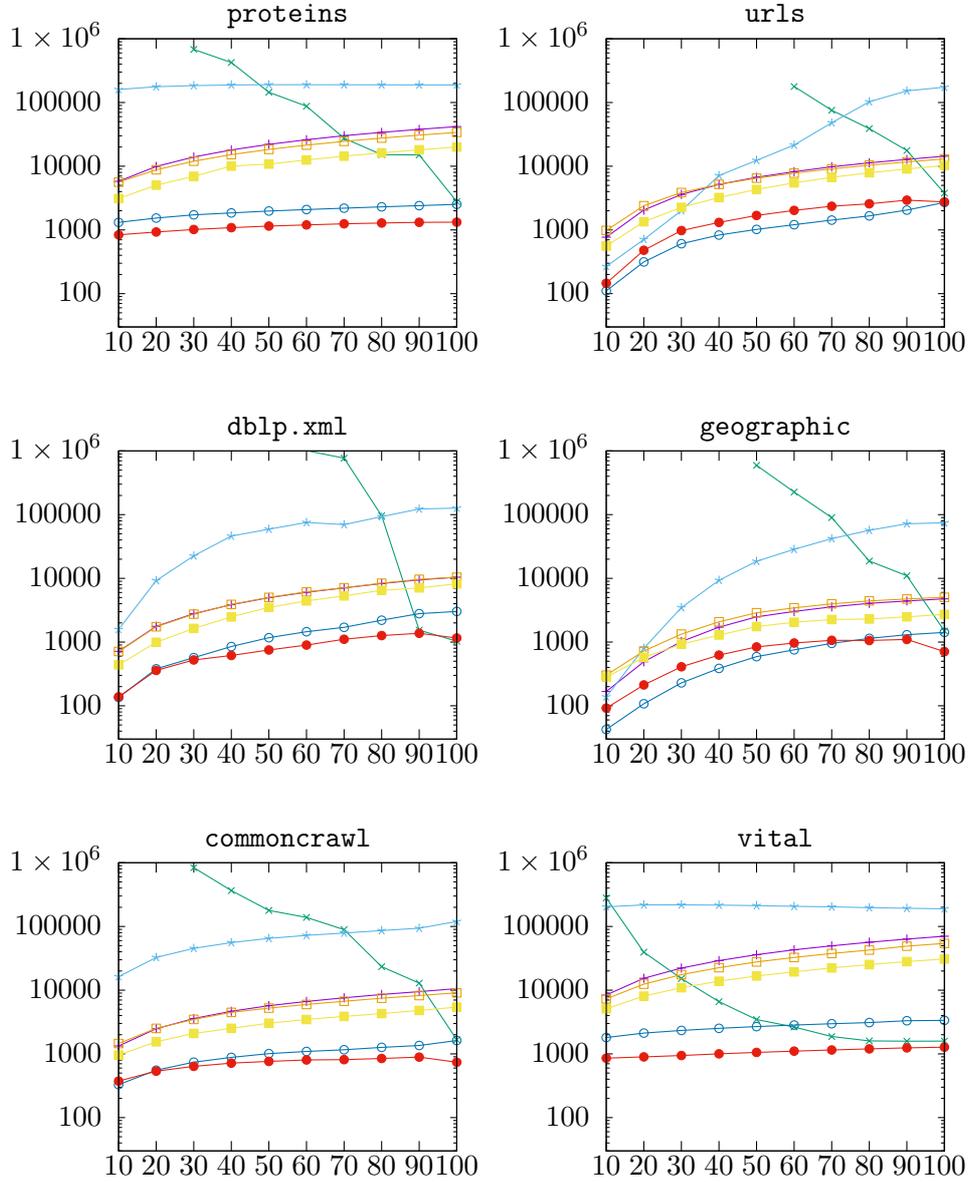

	\begin{minipage}{0.5\linewidth}
	\clearpage{}
\clearpage{}
	\end{minipage}

	\begin{minipage}{0.6\linewidth}
	\caption{Time for answering $\protect\fnLocatePrefix(S)$.
		The y-axis is the average amount of time in nanoseconds (logarithmic scale) for one query.
		The x-axis is the prefix length (in percentage) of the original keyword lengths,
		i.e.,  we search the prefix of length $p |S|/100$ of $S$ at the $x$-axis position $p$ for each keyword~$S$.
	}
	\label{figLocatePrefix}
	\end{minipage}
	\hfill
	\begin{minipage}{0.3\linewidth}
%
 	\end{minipage}
	\vspace{-5em}
\end{figure}

\newcommand*{\WriteHeadRow}{\hline
	\multicolumn{1}{|c}{\setKeys{}}  &
	\multicolumn{1}{|c}{\textsf{CT}} &
	\multicolumn{1}{|c}{\triePCTXOR{}} &
	\multicolumn{1}{|c}{\triePCTHash{}} &
	\multicolumn{1}{|c}{\textsf{ZFT}} &
	\multicolumn{1}{|c}{\TriePP{}} &
	\multicolumn{1}{|c}{} &
	\multicolumn{1}{|c}{\textsf{DA}} &
	\multicolumn{1}{|c|}{$\textsf{HAT-T}$} \\\hline
}

\begin{table*}
		\scriptsize
	\centering
	\subfloat[Time in Nanoseconds]{
		\label{tabInsertTime}
		\begin{tabular}{|l*{5}{|r}|c*{2}{|r}|}
			\WriteHeadRow{}\texttt{proteins   } & 45,508.6 & 45,041.0 & 51,994.3 & 3,683.2 & {\bf 2,349.2} & & 2,088.1 & {\bf 1,805.5}  \\
			\texttt{urls       } & 13,459.0 & 10,580.3 &  8,659.0 & {\bf 4,216.7} & 4,646.1 & & 2,702.8 & {\bf 1,228.9}  \\
			\texttt{dblp.xml   } & 10,066.5 &  8,595.6 &  8,413.1 & 3,309.3 & {\bf 3,035.1} & & {\bf 1,202.8} & 1,371.4  \\
			\texttt{geographic } &  4,711.8 &  4,791.4 &  4,548.5 & {\bf 2,223.4} & 2,427.5 & & 961.6 & {\bf 595.6}  \\
			\texttt{commoncrawl} & 11,077.5 & 11,029.6 & 12,269.6 & 2,368.5 & {\bf 2,260.2} & & 904.9 & {\bf 824.3}  \\
			\texttt{vital      } & 71,666.6 & 75,433.8 & 96,319.3 & 3,515.9 & {\bf 2,002.2} & & {\bf 1,869.5} & 2,151.1  \\
			\hline
		\end{tabular}
	}

	\subfloat[Memory in Megabytes]{
		\label{tabInsertMemory}
		\begin{tabular}{|l*{5}{|r}|c*{2}{|r}|}
			\WriteHeadRow{}\texttt{proteins   } 	& 3,053.12 & 3,053.12 & 4,424.64 & 549.88 & {\bf 418.10} & & 2,142.68 & {\bf 892.66}  \\
			\texttt{urls       } 	& 8,551.47 & 8,551.48 & 9,465.49 & 3,731.14 & {\bf 2,046.61} & & {\bf 932.01} & 1,317.75  \\
			\texttt{dblp.xml   } 	& 1,450.76 & 1,450.77 & 1,871.57 & 552.14 & {\bf 305.74} & & 187.41 & {\bf 144.77}  \\
			\texttt{geographic } 	& 3,029.85 & 3,029.86 & 5,252.34 & 1,204.07 & {\bf 719.36} & & 234.96 & {\bf 164.29}  \\
			\texttt{commoncrawl} 	& 1,040.79 & 1,040.77 & 1,685.80 & 330.03 & {\bf 214.35} & & 269.03 & {\bf 140.81}  \\
			\texttt{vital      } 	&   743.69 &   743.70 & 1,130.68 & 84.29 & {\bf 58.12} & & 322.22 & {\bf 239.12}  \\
			\hline
		\end{tabular}
	}\caption{
		Insertion of all keywords in \emph{random} order.
		We measured (a) the average time per keyword and (b) the memory needed for inserting all keywords of the respective dataset.
		The (a) fastest time and the (b) lowest memory footprint for each keyword set and for each group of trie representations (compact tries or double array tries) are highlighted in bold font.
		For each instance, we measured the maximal virtual memory resident set size (VmRSS),
		which is the second integer in the file \texttt{/proc/self/statm}.
	}
	\label{tabInsert}
\end{table*}

\begin{table}
		\scriptsize
	\centering
	\begin{tabular}{|l*{5}{|r}|c*{2}{|r}|}
		\WriteHeadRow{}\texttt{proteins   } & 42,199.7 & 33,678.0 & 20,011.6 & 2,530.4 & {\bf 1,332.2} & & 1,413.3 & {\bf 609.0}  \\
		\texttt{urls       } & 14,411.8 & 13,087.0 & 10,279.9 & 3,067.1 & {\bf 2,801.6} & & 2,624.1 & {\bf 559.1}  \\
		\texttt{dblp.xml   } & 10,454.9 &  8,990.6 &  6,869.7 & 2,205.5 & {\bf 1,161.4} & & 989.6 & {\bf 439.6}  \\
		\texttt{geographic } &  4,764.3 &  5,016.1 &  2,726.1 & 1,449.5 & {\bf 711.1} & & 423.0 & {\bf 243.6}  \\
		\texttt{commoncrawl} & 10,667.9 &  9,071.7 &  5,423.6 & 1,646.8 & {\bf 742.4} & & 636.8 & {\bf 299.6}  \\
		\texttt{vital      } & 71,552.6 & 52,391.1 & 29,774.7 & 2,806.9 & {\bf 1,204.6} & & 1,138.9 & {\bf 682.6} \\
		\hline
	\end{tabular}
	\caption{
		Average time for $\protect\fnLookup(K)$ in nanoseconds.
		We created a list~$L$ storing all keywords~$K \in \setKeys$, and shuffled it.
		We measured the time of a linear scan over $L$ during which we locate each visited keyword in the respective trie created in  Table~\ref{tabInsert},
		and divided this time by $|\setKeys|$, which yields the average times shown in this table.
	}
	\label{tabLocate}
\end{table}

\subsection{Datasets}
For an objective evaluation, we took a variety of datasets having different characteristics (cf.\ Table~\ref{tabDatasets}):

\begin{itemize}
	\item \texttt{proteins} contains different sequences of amino acids.
	\item \texttt{dblp.xml} is part of the XML dump of the \url{dblp.org} website.
	\item \texttt{urls} is a crawl of webpages of the \texttt{.uk} domain from the WebGraph framework\footnote{\url{http://law.di.unimi.it/webdata/uk-2002}}.
	\item \texttt{geographic} contains names of different geographic locations collected by the GeoNames database\footnote{\url{http://download.geonames.org/export/dump/allCountries.zip}}.
	    Our keywords are extracted from the \texttt{ascii name} column.
	  \item \texttt{commoncrawl} is a web crawl containing the ASCII-encoded content (without HTML tags) of random web pages extracted from Common Crawl\footnote{https://commoncrawl.org/}.
    \item \texttt{vital} is the main text extracted from the most vital Wikipedia articles.
\end{itemize}
The datasets \texttt{proteins} and \texttt{dblp.xml} are from the Pizza\&Chili Corpus\footnote{\url{http://pizzachili.dcc.uchile.cl}}.
The datasets \texttt{commoncrawl} and \texttt{vital} are provided by the tudocomp framework~\cite{dinklage17tudocomp}.

We interpreted each dataset as a single string on the byte alphabet.
We partitioned this string into keywords by splitting it either at newline characters or at full stops, and removed all duplicates afterwards.
The resulting keyword sets are the input of our experiments.

Table~\ref{tabDistributionLengths} lists additional characteristics of the computed keywords of each dataset regarding their lengths and longest common prefixes.
We observe that the lengths have a distribution that is more Gaussian, and by no means uniform.
The lengths have also an impact on the sizes and shapes of the dictionaries, as can be seen in Table~\ref{tabHistogramDict}.

\subsection{Dictionaries of \protect\TriePP{}}
In Table~\ref{tabHistogramDict}, we give insights in how large the dictionaries \DicChild{} and \DicHandle{} of the \TriePP{} become when indexing one of our datasets.
The distributions in Tables~\ref{tabDistributionHandle} and~\ref{tabDistributionChild}
justify our selection of a lightweight data structure with worse asymptotic behavior (FNCS representation) for \DicChild{},
and the use of the more heavyweight cuckoo hash table for \DicHandle{}.
We also did experiments with representing each \DicChild{} as a sorted (or unsorted) list storing newly inserted children with insertion sort (or just at the end of the list).
These experiments showed that lists feature a small speed-up for tiny instances while becoming early slow after a number of insertions, while additionally consuming space for each node, even if they are empty
(remember that represent \DicChild{} of all nodes of a single micro trie \emph{by one} FNCS trie structure.)

\FloatBarrier{}

\subsection{Other Trie Implementations}
We compared \TriePP{} with keyword dictionary representations featuring also a low memory footprint.
We present two groups of trie implementations.
The first group consists of two non-compact trie data structures:

\begin{itemize}
    \item \textsf{DA}: the double array~\cite{aoe89doublearray} implementation of the Cedar library\footnote{\url{http://www.tkl.iis.u-tokyo.ac.jp/~ynaga/cedar/}}. \item \textsf{HAT-T}: the HAT-trie~\cite{askitis10trie} implementation of Tessil\footnote{\url{https://github.com/Tessil/hat-trie}}.
Tessil's implementation exploits that keywords have a small length in practice.
The default implementation assumes that all these lengths can be stored in 16 bits, which is not true for the dataset \texttt{commoncrawl}.
We therefore evaluated the HAT-trie with 16 and 32 bits for the lengths, and took the minimum time and minimum space of both variants throughout the evaluation.
\end{itemize}
As we will see in the following,
the keyword dictionaries of the first group are
lightweight and overall efficient but perform prefix searches poorly.
The second group consists of other compact trie data structures:

\begin{itemize}
    \item \textsf{CT}: a compact trie without word packing.
	\item \triePCTXOR{}: a packed c-trie using bit parallelism to compare compact words.
	\item \triePCTHash{}: a packed c-trie using additionally the hash table implementation \texttt{unordered\_map} of the \CPlusPlus{} standard library as a dictionary in each micro trie for retrieving a node by its extent (it is similar to our \DicHandle{}, but uses the extents instead of the handles as keys).
	\item \textsf{ZFT}: our z-fast trie portation from an implementation in Java\footnote{This implementation is part of Vigna's Sux4J library, located at \url{https://github.com/vigna/Sux4J}.} to \CPlusPlus{}.
		We added an evaluation of the original Java version to the appendix.
\end{itemize}
The implementations of the compact trie and the packed c-tries are due to Takagi et al.~\cite{takagi17packedtrie}.
The implementations \triePCTXOR{} and \triePCTHash{} pack characters in 32-bit integers,
whereas all other implementations use 64-bit integers, which reflect the machine word size of commodity computers nowadays.
All implementations (of both groups) are written in \CPlusPlus{}, and compiled with \texttt{gcc-8.2.0} in the highest optimization mode \texttt{-O3}.

In what follows, we evaluate these trie implementations on the aforementioned datasets.

\subsection{Construction}
In the first experiment, we measured the time it takes to insert all keywords of a dataset into a keyword dictionary in random order.
We give the results in Table~\ref{tabInsert}.
This table reveals that the construction of \TriePP{} is faster than the construction of every packed trie (i.e., \textsf{CT}, \triePCTXOR{}, \triePCTHash{}, and \textsf{ZFT}).
Except for \textsf{ZFT}, its final size is also an improvement to the sizes of those data structures.
If the average keyword length is sufficiently large, \TriePP{} is memory-friendlier than \textsf{DA} and \textsf{HAT-T} (e.g.\ \texttt{proteins} or \texttt{vital}) while
it is inferior in both space and time when maintaining mostly short keywords.

\subsection{Locate Prefix Queries}

A major highlight is the time needed for $\fnLocatePrefix(S)$ queries shown in Fig.~\ref{figLocatePrefix}.
Instead of returning an iterator to a set as requested at the beginning of this article,
we require each keyword dictionary to return the complete set of all keywords having~$S$ as a prefix.
In this setting, \TriePP{} dominates most of the time.
We observe that \textsf{DA} becomes faster for longer prefixes.
This effect can be explained as follows:
First recognize by Table~\ref{tabLocate} that \textsf{DA} has competitive \fnLookup{} times, allowing the trie to match a pattern at high speed.
The matching locates the lowest node~$v$ whose extent is a prefix of $S$.
After locating~$v$, it resorts to exploring the entire subtree of~$v$, 
which is a slow operation for large subtrees.
If~$v$ is a deep node, chances are that its subtree size is rather small, enabling \textsf{DA} to process $v$'s subtree quickly.

\subsection{Lookup Queries}
The results for \fnLookup{} are collected in Table~\ref{tabLocate}.
In all instances, \TriePP{} answered \fnLookup{} queries faster than all packed tries.
However, \textsf{HAT-T}, followed by \textsf{DA}, provide the fastest solutions for answering \fnLookup{}.

\newcommand*{\WriteHeadRowDelete}{\hline
	\multicolumn{1}{|c}{\setKeys{}}  &
	\multicolumn{1}{|c}{\textsf{ZFT}} &
	\multicolumn{1}{|c}{\TriePP{}} &
	\multicolumn{1}{|c}{} &
	\multicolumn{1}{|c}{\textsf{DA}} &
	\multicolumn{1}{|c|}{$\textsf{HAT-T}$} \\\hline
}

\begin{table*}
	\centering
		\begin{tabular}{|l*{5}{|r}|c*{2}{|r}|}
			\WriteHeadRowDelete{}\texttt{proteins   } & 3,676.4 & {\bf 2,012.0} & & 1,606.1 & {\bf 1,187.7}  \\
			\texttt{urls       } & 5,677.7 & {\bf 4,045.5} & & 3,060.8 & {\bf 886.5}  \\
			\texttt{dblp.xml   } & 3,501.5 & {\bf 2,219.4} & & 1,211.7 & {\bf 667.4}  \\
			\texttt{geographic } & 2,254.6 & {\bf 1,761.8} & & 787.8 & {\bf 494.3}  \\
			\texttt{commoncrawl} & 2,526.7 & {\bf 1,645.4} & & 868.5 & {\bf 573.1}  \\
			\texttt{vital      } & 3,727.4 & {\bf 1,780.8} & & {\bf 1,042.0} & 1,302.7  \\
			\hline
		\end{tabular}

		\caption{Average time for $\protect\fnDelete(K)$ in nanoseconds.}
    \label{tabDelete}
\end{table*}

\subsection{Deletions}
We also ran experiments for the delete operation, which we conducted in the same fashion as the experiments for \fnLookup{}.
We put the results in Table~\ref{tabDelete}.
There, we omit the implementations for \textsf{CT}, \triePCTXOR{}, and \triePCTHash{} since they do not provide a delete operation.
We observe that \TriePP{} is always faster than \textsf{ZFT}, but at most 3 times slower than \textsf{DA}, and 2 to 5 times slower than \textsf{HAT-T}.

\begin{table*}
		\scriptsize
	\centering
    \subfloat[Time in Nanoseconds]{\label{tabInsertSortedTime}
		\begin{tabular}{|l*{5}{|r}|c*{2}{|r}|}
			\WriteHeadRow{}\texttt{proteins   } & 39,716.6 & 38,547.4 & 48,384.0 & 2,623.4 & {\bf 1,369.1} & & 1,225.3 & {\bf 853.3}  \\
			\texttt{urls       } &  9,849.2 &  6,398.6 &  4,786.9 & 2,604.1 & {\bf 709.5} & & {\bf 610.9} & 480.7  \\
			\texttt{dblp.xml   } &  7,736.4 &  5,713.0 &  5,645.8 & 2,051.3 & {\bf 736.6} & & {\bf 451.9} & 810.4  \\
			\texttt{geographic } &  2,342.1 &  2,089.6 &  2,605.7 & 1,305.1 & {\bf 1,035.8} & & {\bf 237.0} & 258.3  \\
			\texttt{commoncrawl} &  8,419.2 &  8,012.2 &  9,930.3 & 1,485.4 & {\bf 1,072.3} & & {\bf 370.3} & 385.4  \\
			\texttt{vital      } & 63,719.1 & 65,684.8 & 90,066.2 & 3,187.2 & {\bf 865.9} & & 1,313.2 & {\bf 1,266.6}  \\
			\hline
		\end{tabular}
	}

	\subfloat[Memory in Megabytes]{\label{tabInsertSortedMemory}
		\begin{tabular}{|l*{5}{|r}|c*{2}{|r}|}
			\WriteHeadRow{}\texttt{proteins 	 } & 2,889.31 & 2,889.32 &  4,376.2 & 549.87 & {\bf 422.68} & & 1,779.47 & {\bf 890.14}  \\
			\texttt{urls       } & 8,533.40 & 8,533.41 & 10,027.4 & 3,731.14 & {\bf 2,046.45} & & {\bf 1,017.18} & 1,302.21  \\
			\texttt{dblp.xml 	 } & 1,445.39 & 1,445.40 &  1,850.2 & 552.14 & {\bf 305.70} & & 173.62 & {\bf 141.59}  \\
			\texttt{geographic } & 3,029.50 & 3,029.51 &  4,952.8 & 1,204.07 & {\bf 719.35} & & 251.86 & {\bf 159.23}  \\
			\texttt{commoncrawl} & 1,023.88 & 1,023.87 &  1,598.8 & 330.03 & {\bf 220.18} & & 174.45 & {\bf 139.61}  \\
			\texttt{vital      } &   695.96 &   695.97 &  1,098.4 & 84.29 & {\bf 58.12} & & 261.09 & {\bf 238.09}  \\
			\hline
		\end{tabular}
	}\caption{Insertion of all keywords in \emph{lexicographical} order.
		Except to the ordering of the keywords, the setting is the same as in Table~\ref{tabInsert}.
	}
	\label{tabInsertSorted}
\end{table*}

\subsection{Sorted Insertions}
Up to now, we covered the case of creating a trie on keywords shuffled in a random order~$R$,
and subsequently queried the trie with the keywords in another random order~$R'$.
However, one might question whether other possibilities like building a keyword dictionary with lexicographically sorted keywords, or querying it with keywords arranged in the same order as in the construction is advantageous.
For that, we revisit the construction in Table~\ref{tabInsertSorted}, filling a keyword dictionary now with keywords in lexicographically sorted order.
Compared to Table~\ref{tabInsert},
the space requirement in both scenarios is nearly the same for each keyword dictionary.
However, a lexicographically sorted insertion speeds up the construction of all of instances.
Especially the construction times of \TriePP{} seem to be order sensitive, 
as we now observe a large gap in the running times between \TriePP{} and \textsf{ZFT}.
In the sorted insertion order, 
\TriePP{} is always the best among all packed trie representations, and for \texttt{vital}, it is even the overall best representation.

\begin{table*}
		\scriptsize
	\centering
    \subfloat[Sorted - Sorted]{\label{tabLocateSorted}
\begin{tabular}{|l*{5}{|r}|c*{2}{|r}|}
			\WriteHeadRow{}\texttt{proteins   } & 39,357.5 & 30,758.8 & 18,392.0 & 1,748.6 & {\bf 421.1} & & 391.6 & {\bf 256.0}  \\
			\texttt{URLs       } & 10,296.2 &  8,781.5 &  5,945.6 & 1,223.0 & {\bf 240.6} & & 155.2 & {\bf 138.7}  \\
			\texttt{dblp.xml   } &  7,957.8 &  6,372.7 &  4,481.0 & 1,121.3 & {\bf 190.4} & & {\bf 111.9} & 136.0  \\
			\texttt{geographic } &  1,839.2 &  1,925.0 &  1,436.4 & 717.3 & {\bf 179.6} & & {\bf 44.8} & 65.4  \\
			\texttt{commoncrawl} &  8,273.8 &  6,555.0 &  4,294.6 & 930.4 & {\bf 169.7} & & {\bf 95.9} & 99.6  \\
			\texttt{vital      } & 69,059.5 & 49,871.5 & 28,850.4 & 2,046.7 & {\bf 404.6} & & 526.3 & {\bf 346.1}  \\
			\hline
		\end{tabular}
	}

	\subfloat[Sorted - Order~$R$]{\label{tabLocateSortedShuffled}
		\begin{tabular}{|l*{5}{|r}|c*{2}{|r}|}
			\WriteHeadRow{}\texttt{proteins   } & 40,154.2 & 31,487.0 & 18,817.7 & 2,440.1 & {\bf 1,155.6} & & 1,084.5 & {\bf 627.3}  \\
			\texttt{urls       } & 10,725.0 & 9,287.6 & 6,376.8 & {\bf 2,476.8} & 2,470.9 & & 1,739.7 & {\bf 575.4}  \\
			\texttt{dblp.xml   } & 8,194.8 & 6,609.8 & 4,781.5 & 2,054.1 & {\bf 1,084.9} & & 746.0 & {\bf 459.7}  \\
			\texttt{geographic } & 2,054.8 & 2,130.7 & 1,376.0 & 1,288.4 & {\bf 636.4} & & 353.5 & {\bf 246.9}  \\
			\texttt{commoncrawl} & 8,697.6 & 6,892.4 & 4,220.5 & 1,575.9 & {\bf 627.5} & & 470.1 & {\bf 305.6}  \\
			\texttt{vital      } & 71,081.0 & 53,701.4 & 29,366.9 & 2,726.1 & {\bf 1,111.8} & & 1,020.7 & {\bf 681.9}  \\
			\hline
		\end{tabular}
	}

	\subfloat[Order~$R$ - Sorted]{\label{tabLocateShuffledSorted}
		\begin{tabular}{|l*{5}{|r}|c*{2}{|r}|}
			\WriteHeadRow{}\texttt{proteins   } & 42,934.5 & 33,231.1 & 19,988.6 & 2,050.3 & {\bf 805.6} & & 691.6 & {\bf 309.3}  \\
			\texttt{urls       } & 14,563.1 & 12,500.3 & 9,321.5 & 1,598.1 & {\bf 635.6} & & 361.4 & {\bf 177.5}  \\
			\texttt{dblp.xml   } & 10,180.9 & 8,702.2 & 6,496.8 & 1,451.4 & {\bf 473.0} & & 297.9 & {\bf 161.5}  \\
			\texttt{geographic } & 4,408.0 & 4,665.0 & 3,746.0 & 959.1 & {\bf 407.6} & & 162.3 & {\bf 84.9}  \\
			\texttt{commoncrawl} & 10,370.6 & 8,761.5 & 6,016.1 & 1,175.2 & {\bf 423.4} & & 267.0 & {\bf 123.3}  \\
			\texttt{vital      } & 71,992.7 & 53,526.8 & 30,583.7 & 2,341.1 & {\bf 778.6} & & 788.0 & {\bf 411.4}  \\
			\hline
		\end{tabular}
	}

	\subfloat[Order $R$ - Order $R$]{\label{tabLocateShuffledShuffled}
		\begin{tabular}{|l*{5}{|r}|c*{2}{|r}|}
			\WriteHeadRow{}\texttt{proteins   } & 42,134.8 & 33,626.8 & 19,904.1 & 2,299.5 & {\bf 1,016.8} & & 1,211.9 & {\bf 605.4}  \\
			\texttt{urls       } & 14,329.6 & 13,008.3 & 10,187.0 & 2,462.7 & {\bf 2,410.2} & & 2,491.6 & {\bf 556.6}  \\
			\texttt{dblp.xml   } & 10,398.2 & 8,938.6 & 6,801.3 & 1,979.3 & {\bf 920.32} & & 863.5 & {\bf 436.3}  \\
			\texttt{geographic } & 4,703.1 & 4,966.8 & 2,644.0 & 1,296.5 & {\bf 501.7} & & 387.6 & {\bf 240.4}  \\
			\texttt{commoncrawl} & 10,624.8 & 9,040.9 & 5,353.6 & 1,496.8 & {\bf 550.5} & & 553.6 & {\bf 295.7}  \\
			\texttt{vital      } & 71,523.2 & 52,342.0 & 29,680.5 & 2,579.2 & {\bf 943.7} & & 952.8 & {\bf 665.1}  \\
			\hline
		\end{tabular}
	}\caption{Average time for $\protect\fnLookup(K)$ in nanoseconds.
		We create a trie by inserting keywords contained a list~$L$ whose elements are (a-b) lexicographically sorted or (c-d) in a random order~$R$.
		We stick to the setting of Table~\ref{tabLocateComplement}, where we used~$L$ for the queries.
		However, before the querying, we (b) shuffled~$L$, (a,c) sorted the elements in~$L$ lexicographically, or (d) kept~$L$ as it is.
    }
    \label{tabLocateComplement}
\end{table*}

\begin{figure*}
    \begin{minipage}{0.5\linewidth}
		\clearpage{}%
\clearpage{}
    \end{minipage}

    \begin{minipage}{0.6\linewidth}
		\caption{Time for answering $\protect\fnLocatePrefix$ when the data structures are built and queried with the keywords in lexicographical sorted order.
		The setting is, except from the different order, the same as in Fig.~\ref{figLocatePrefix}.}
		\label{figLocatePrefixSorted}
    \end{minipage}
    \hfill
    \begin{minipage}{0.3\linewidth}
		%
\clearpage{}
	\end{minipage}

	\begin{minipage}{0.6\linewidth}
		\caption{Time for answering $\protect\fnLocatePrefix$ when the data structures are built with the keywords in lexicographical sorted order, but queried with the keywords in random order.
		The setting is, except from the different orders, the same as in Fig.~\ref{figLocatePrefix}.}
		\label{figLocatePrefixSortedShuffled}
	\end{minipage}
	\hfill
	\begin{minipage}{0.3\linewidth}
		%
\clearpage{}
	\end{minipage}

	\begin{minipage}{0.6\linewidth}
		\caption{Time for answering $\protect\fnLocatePrefix$ when the data structures are built with the keywords in random order, but queried with the keywords sorted in lexicographical order.
			The setting is, except from the different orders, the same as in Fig.~\ref{figLocatePrefix}.}
		\label{figLocatePrefixShuffledSorted}
	\end{minipage}
	\hfill
	\begin{minipage}{0.3\linewidth}
		%
\clearpage{}
	\end{minipage}

	\begin{minipage}{0.6\linewidth}
		\caption{Time for answering $\protect\fnLocatePrefix$ when the data structures are built with the keywords in a random order~$O$, and queried with the keywords in the same order~$O$.
			The setting is, except from the different order, the same as in Fig.~\ref{figLocatePrefix}.}
		\label{figLocatePrefixShuffledShuffled}
	\end{minipage}
	\hfill
	\begin{minipage}{0.3\linewidth}
		
 	\end{minipage}
\end{figure*}

\begin{table*}
	\centering

    \subfloat[Sorted - Sorted]{\label{tabDeleteSorted}
		\begin{tabular}{|l*{5}{|r}|c*{2}{|r}|}
			\WriteHeadRowDelete{}\texttt{proteins   } & 2,457.1 & {\bf 875.4} & & {\bf 476.1} & 568.3  \\
			\texttt{urls       } & 2,471.2 & {\bf 526.4} & & {\bf 211.6} & 293.1  \\
			\texttt{dblp.xml   } & 2,059.1 & {\bf 520.6} & & {\bf 153.3} & 236.3  \\
			\texttt{geographic } & 1,161.2 & {\bf 608.1} & & {\bf 77.6} & 143.4  \\
			\texttt{commoncrawl} & 1,465.1 & {\bf 575.0} & & {\bf 129.5} & 207.6  \\
			\texttt{vital      } & 2,704.5 & {\bf 813.3} & & {\bf 437.0} & 733.8  \\
			\hline
		\end{tabular}
	}

	\subfloat[Sorted - Order~$R$]{\label{tabDeleteSortedShuffled}
		\begin{tabular}{|l*{5}{|r}|c*{2}{|r}|}
			\WriteHeadRowDelete{}\texttt{proteins   } & 2,455.5 & {\bf 874.1} & & {\bf 476.0} & 567.4  \\
			\texttt{urls       } & 2,476.7 & {\bf 525.6} & & {\bf 211.5} & 292.6  \\
			\texttt{dblp.xml   } & 2,065.0 & {\bf 522.4} & & {\bf 153.4} & 237.1  \\
			\texttt{geographic } & 1,158.1 & {\bf 608.3} & & {\bf 77.6}  & 144.1  \\
			\texttt{commoncrawl} & 1,561.8 & {\bf 574.5} & & {\bf 129.5} & 207.5  \\
			\texttt{vital      } & 2,692.4 & {\bf 814.1} & & {\bf 435.7} & 764.0  \\
			\hline
		\end{tabular}
	}

	\subfloat[Order~$R$ - Sorted]{\label{tabDeleteShuffledSorted}
		\begin{tabular}{|l*{5}{|r}|c*{2}{|r}|}
			\WriteHeadRowDelete{}\texttt{proteins   } & 3,651.9 & {\bf 2,012.4} & & 1,603.5 & {\bf 1,187.7}  \\
			\texttt{urls       } & 4,257.0 & {\bf 3,976.2} & & 3,051.1 & {\bf 883.9}  \\
			\texttt{dblp.xml   } & 3,493.1 & {\bf 2,141.1} & & 1,221.0 & {\bf 671.1}  \\
			\texttt{geographic } & 2,296.8 & {\bf 1,750.3} & & 782.9 & {\bf 494.6}  \\
			\texttt{commoncrawl} & 2,513.5 & {\bf 1,632.9} & & 875.8 & {\bf 568.9}  \\
			\texttt{vital      } & 3,701.9 & {\bf 1,781.7} & & {\bf 1,026.0} & 1,305.5  \\
			\hline
		\end{tabular}
	}

	\subfloat[Order $R$ - Order $R$]{\label{tabDeleteShuffledShuffled}
		\begin{tabular}{|l*{5}{|r}|c*{2}{|r}|}
			\WriteHeadRowDelete{}\texttt{proteins   } & 3,959.9 & {\bf 2,006.1} & & 1,607.1 & {\bf 1,194.2}  \\
			\texttt{urls       } & 4,340.2 & {\bf 3,986.6} & & 3,061.0 & {\bf 885.7}  \\
			\texttt{dblp.xml   } & 3,438.7 & {\bf 2,147.2} & & 1,227.4 & {\bf 668.4}  \\
			\texttt{geographic } & 2,236.0 & {\bf 1,765.64} & & 783.1 & {\bf 493.7}  \\
			\texttt{commoncrawl} & 2,516.1 & {\bf 1,629.7} & & 878.6 & {\bf 570.9}  \\
			\texttt{vital      } & 3,696.0 & {\bf 1,779.0} & & {\bf 1,041.6} & 1,297.9 \\
			\hline
		\end{tabular}
	}\caption{Average time for $\protect\fnDelete(K)$ in nanoseconds under different orders.
		The setting with two different random orders~$R$ and~$R'$ (Order $R$ - Order $R'$) is already presented in Table~\ref{tabDelete}, which has the same setting as Table~\ref{tabLocate}.
		For the other sub-tables, the setting is given in Table~\ref{tabLocateComplement}.
    }
    \label{tabDeleteSorting}
\end{table*}

\FloatBarrier{}

\subsection{Order of Queries}
Having two scenarios for trie construction, we can also think about different orders of how to query the data structures.
Here, we present a Cartesian product of these orders, shown in Table~\ref{tabLocateComplement} for \fnLookup{},
and in Figs.~\ref{figLocatePrefixSorted}, \ref{figLocatePrefixSortedShuffled}, \ref{figLocatePrefixShuffledSorted}, and \ref{figLocatePrefixShuffledShuffled} for \fnLocatePrefix{}.
We see a remarkable speedup of the query operations of all keyword dictionary implementations
when they are fed with keywords in lexicographically order.
The best bets can be placed on the setting of Table~\ref{tabLocateSorted} and Fig.~\ref{figLocatePrefixSorted},
where especially \TriePP{} shows a performance boost, being competitive to \textsf{DA} on some dataset instances for \fnLookup{} (\texttt{proteins} or \texttt{vital}).
A slightly slower variant is to query in random order (Table~\ref{tabLocateSortedShuffled} and Fig.~\ref{figLocatePrefixSortedShuffled}).
The execution times of the keyword dictionaries fed in random order follow with a large gap.
Here, the order in which the queries are executed has also an impact on the execution times, but is not as large as we have seen for the case where we inserted the keywords in lexicographic order.
We obtain the fastest execution times when querying the keywords in lexicographic order
(Table~\ref{tabLocateShuffledSorted} and Fig.~\ref{figLocatePrefixShuffledSorted}).
Regarding the query order, the compact trie and the packed c-tries have roughly the same query times for different orders, 
unlike other trie data structure having a noticeable speed-up.
Especially \textsf{ZFT} and \TriePP{} can take advantage of the case when the queries are in lexicographic order (Table~\ref{tabLocateShuffledSorted} and Fig.~\ref{figLocatePrefixShuffledSorted}).
Finally, Table~\ref{tabDeleteSorting} complements Table~\ref{tabDelete} for the study of the delete operations regarding different orders,
where the setting of Table~\ref{tabDelete} can be interpreted as inserting the keywords and querying all keywords in two different random orders~$R$ and~$R'$, respectively.
Here, we observe similar characteristics to the study of \fnLookup{},
with the difference that the performance gap between \textsf{DA} and \TriePP{} is unfortunately larger.

\section{Conclusion}
We have presented the trie data structure \TriePP{} to cope with the demands for fast prefix searches
like auto-completion~\cite{cai16autocompletion}.
In case that prefix queries dominate dynamic operations like insertions with respect to their quantity,
the keyword dictionary \TriePP{} offers one of the best trade-offs among all tested candidates.

For future work, we can speed up the insertions of keywords that share long prefixes with other keywords by vectorization.
That is because the word packing approach for comparing two strings interpreted as two packed strings can be vectorized.
Recent instruction sets like AVX feature instructions for this task.
An application\footnote{\url{https://github.com/koeppl/packed_string}} shows that the computation time roughly halves for long enough common prefixes when exploiting the AVX2 instruction set.

Table~\ref{tabDistributionHandle} reveals that some instances of \DicHandle{} grow extremely large while most of the other instances maintain only few entries.
For the large ones, we can use a compact hash table such as~\cite{koppl20hash}\footnote{We target to change this hash table to a bucketized cuckoo hash table like~\cite{dietzfelbinger07cuckoo}.
} 
that stores quotients instead of the values, 
where a quotient has bit length $v - \lg M$ 
if the values can be represented in $v$ bits (we set $v$ to 32 bits in Sect.~\ref{secImplementation}),
where $M$ is the number of cells of the hash table.

Considering different hash table layouts, we conducted an experiment with the linear probing hash table of Rigtorp\footnote{\url{https://github.com/rigtorp/HashMap}} storing nodes along with the (redundant) keys.
While using considerably more space, 
this hash table performed only slightly better than the cuckoo hash table, 
even when storing the keys explicitly and with a load factor of $0.5$.
Dropping the keys as we did in Sect.~\ref{secImplementation}, a hash table with linear probing will likely be outperformed by our cuckoo hash table as cache effects become negligible.

As stated in the caption of Table~\ref{tabHistogramDict},
none of our datasets is prefix-free. 
In a more enhanced evaluation, we would like to conduct our experiments after a preprocessing step in which we discard every keyword that is a prefix of another keyword.

\paragraph{Funding: }
This work was supported by JSPS KAKENHI Grant Numbers JP18F18120 (DK), JP18K18002 (YN), JP17H01697 (SI),
JP16H02783 (HB), JP20H04141 (HB), JP18H04098 (MT), and by JST PRESTO Grant Number JPMJPR1922 (SI).

\bibliographystyle{abbrv}
\bibliography{literature,refs}

\clearpage
\appendix

\section{Original z-fast trie}
The original implementation of the z-fast trie of Vigna is written in Java as part of his Sux4J library.
As a supplement, we conducted our experiments of this implementation on the same machine.
However, we could not build this trie for the keyword set \texttt{vital}.
The time and space needed for the trie construction are given in Table~\ref{tabInsertZFT}.
Its time for \fnLookup{} and \fnLocatePrefix{} are shown in Table~\ref{tabLocateZFT} and Fig.~\ref{figLocatePrefixZFT}, respectively.
Its time for \fnDelete{} is given in Table~\ref{tabDeleteZFT}.
Unfortunately, we received runtime failures on several instances, which we marked with \emph{N/A} (for not available) in the experiments.

\begin{table*}
	\centering
	\subfloat[Time in Nanoseconds]{\label{tabZFTInsertTime}
\begin{tabular}{l*{2}{|r}}
			\hline
			\multicolumn{1}{c}{\setKeys{}}  &
			\multicolumn{1}{|c}{Random} &
			\multicolumn{1}{|c}{Sorted} \\
			\hline \texttt{proteins   } & 3,896.6 & 2,764.8 \\
			\texttt{urls       } & 3,056.7 & 2,038.7 \\
			\texttt{dblp.xml   } & 2,727.1 & 1,693.6 \\
			\texttt{geographic } & 2,802.9 & 1,831.0 \\
			\texttt{commoncrawl} & 2,883.3 & 1,714.9 \\
			\texttt{vital      } & N/A & N/A \\
			\hline
		\end{tabular}
	}\hspace{1cm}
	\subfloat[Memory in Megabytes]{\label{tabZFTInsertMemory}
		\begin{tabular}{l*{2}{|r}}
			\hline
			\multicolumn{1}{c}{\setKeys{}}  &
			\multicolumn{1}{|c}{Random} &
			\multicolumn{1}{|c}{Sorted} \\
			\hline \texttt{proteins   } & 1,629.60 & 1,630.81 \\
			\texttt{urls       } & 2,764.73 & 2,341.69 \\
			\texttt{dblp.xml   } &   989.38 & 1,026.62 \\
			\texttt{geographic } & 1,043.65 & 1,075.91 \\
			\texttt{commoncrawl} &   244.94 &   245.73 \\
			\texttt{vital      } & N/A    & N/A \\
			\hline
		\end{tabular}
	}

	\caption{Inserting of all keywords in the z-fast trie Java-implementation.}
	\label{tabInsertZFT}
\end{table*}

\begin{table*}
	\centering
	\begin{tabular}{l*{5}{|r}}
		\hline
		\multicolumn{1}{c}{\setKeys{}}  &
		\multicolumn{1}{|c}{R-R'} &
		\multicolumn{1}{|c}{S-S} &
		\multicolumn{1}{|c}{S-R} &
		\multicolumn{1}{|c}{R-S} &
		\multicolumn{1}{|c}{R-R} \\
		\hline \texttt{proteins   } & 5,093.5 & 4,798.0 & 5,403.4 & 5,136.4 & 5,052.2 \\
		\texttt{urls       } & 2,384.2 & 1,655.2 & 2,615.3 & 1,730.9 & 2,438.1 \\
		\texttt{dblp.xml   } & 1,778.6 & 1,322.2 & 1,848.7 & 1,265.9 & 2,165.1 \\
		\texttt{geographic } & 1,254.7 & 749.2 & 1,416.0 & 1,154.6 & 1,233.8 \\
		\texttt{commoncrawl} & 2,032.3 & 1,351.9 & 1,870.8 & 1,404.8 & 1,648.0 \\
		\texttt{vital      } & N/A & N/A & N/A & N/A & N/A \\
		\hline
	\end{tabular}
	\caption{Average time for answering $\protect\fnLookup(K)$ with the z-fast trie Java-implementation. Times are in nanoseconds.
		The table covers the settings of Tables~\ref{tabLocate} (Order $R$ - Order $R'$),
		\ref{tabLocateSorted} (Sorted - Sorted),
		\ref{tabLocateShuffledSorted} (Order $R$ - Sorted),
		\ref{tabLocateSortedShuffled} (Sorted - Order $R$), and
		\ref{tabLocateShuffledShuffled} (Order $R$ - Order $R$), where $R$ and $R'$ are two different random orderings.
	}
	\label{tabLocateZFT}
\end{table*}

\begin{table*}
	\centering
	\begin{tabular}{l*{5}{|r}}
		\hline
		\multicolumn{1}{c}{\setKeys{}}  &
		\multicolumn{1}{|c}{R-R'} &
		\multicolumn{1}{|c}{S-S} &
		\multicolumn{1}{|c}{S-R} &
		\multicolumn{1}{|c}{R-S} &
		\multicolumn{1}{|c}{R-R} \\
		\hline \texttt{proteins   } & 5358.4 & 4173.3 & 4899.6 & 4205.4 & 4421.9 \\
		\texttt{urls       } & 4412.0 & 2904.0 & 3997.9 & 2848.4 & 4253.2 \\
		\texttt{dblp.xml   } & N/A & N/A & N/A & N/A & N/A \\
		\texttt{geographic } & 2476.4 & 1223.1 & 1525.9 & 2690.6 & 2344.4 \\
		\texttt{commoncrawl} & N/A & N/A & N/A & N/A & N/A \\
		\texttt{vital      } & N/A & N/A & N/A & N/A & N/A \\
		\hline
	\end{tabular}
	\caption{Average time for answering $\protect\fnDelete(K)$ with the z-fast trie Java-implementation.
		The meaninng of the column captions is the same as in Table~\ref{tabLocateZFT}.
	}
	\label{tabDeleteZFT}
\end{table*}

\begin{figure*}
	\begin{minipage}{0.5\linewidth}
		\clearpage{}%
\clearpage{}
	\end{minipage}

	\begin{minipage}{0.6\linewidth}
		\caption{Time for answering $\protect\fnLocatePrefix$ with the z-fast trie Java-implementation.
			The plots cover the settings of Figs.~\ref{figLocatePrefix} (Order $R$ - Order $R'$),
			\ref{figLocatePrefixSorted} (Sorted - Sorted),
			\ref{figLocatePrefixSortedShuffled} (Order $R$ - Sorted),
			\ref{figLocatePrefixShuffledSorted} (Sorted - Order $R$), and
			\ref{figLocatePrefixShuffledShuffled} (Order $R$ - Order $R$), where $R$ and $R'$ are two different random orderings.
		}
		\label{figLocatePrefixZFT}
	\end{minipage}
	\hfill
	\begin{minipage}{0.3\linewidth}
		\begin{tikzpicture}[gnuplot]
\gpcolor{color=gp lt color border}
\node[gp node right] at (9.979,6.280) {R-R'};
\gpcolor{rgb color={0.580,0.000,0.827}}
\gpsetlinetype{gp lt border}
\gpsetdashtype{gp dt solid}
\gpsetlinewidth{1.00}
\draw[gp path] (10.163,6.280)--(11.079,6.280);
\gpsetpointsize{4.00}
\gppoint{gp mark 1}{(10.621,6.280)}
\gpcolor{color=gp lt color border}
\node[gp node right] at (9.979,5.818) {S-S};
\gpcolor{rgb color={0.000,0.620,0.451}}
\draw[gp path] (10.163,5.818)--(11.079,5.818);
\gppoint{gp mark 2}{(10.621,5.818)}
\gpcolor{color=gp lt color border}
\node[gp node right] at (9.979,5.356) {R-S};
\gpcolor{rgb color={0.337,0.706,0.914}}
\draw[gp path] (10.163,5.356)--(11.079,5.356);
\gppoint{gp mark 3}{(10.621,5.356)}
\gpcolor{color=gp lt color border}
\node[gp node right] at (9.979,4.894) {S-R};
\gpcolor{rgb color={0.902,0.624,0.000}}
\draw[gp path] (10.163,4.894)--(11.079,4.894);
\gppoint{gp mark 4}{(10.621,4.894)}
\gpcolor{color=gp lt color border}
\node[gp node right] at (9.979,4.432) {R-R};
\gpcolor{rgb color={0.941,0.894,0.259}}
\draw[gp path] (10.163,4.432)--(11.079,4.432);
\gppoint{gp mark 5}{(10.621,4.432)}
\gpdefrectangularnode{gp plot 1}{\pgfpoint{0.246cm}{0.368cm}}{\pgfpoint{11.447cm}{6.691cm}}
\end{tikzpicture}
 	\end{minipage}
\end{figure*}


\end{document}